\documentclass[a4paper,11pt]{article}
\pdfoutput=1 % if your are submitting a pdflatex (i.e. if you have
\usepackage{jheppub} % for details on the use of the package, please see the JINST-author-manual
\usepackage{lineno}
\usepackage{slashed}
\usepackage{subcaption}
\usepackage{mathtools}
\usepackage{commath}
\usepackage{multirow}
\usepackage{hyperref}
\usepackage{cleveref}
\usepackage{booktabs}
\usepackage{nicefrac} 
\usepackage{comment}
\usepackage{graphicx}
\usepackage{float}
\usepackage{enumitem}
\usepackage{physics}

\title{Vacuum Instability and False Vacuum Decay Induced by Domain Walls in the N2HDM}

\author[a]{Mohamed Younes Sassi}
\author[a,b]{Gudrid Moortgat-Pick}

\affiliation[a]{II. Institut für Theoretische Physik,
University of Hamburg,\\Luruper Chaussee 149, 22761 Hamburg, Germany}
\affiliation[b]{Deutsches Elektronen-Synchrotron DESY, Notkestr. 85, 22607 Hamburg, Germany}

\preprint{DESY-25-084}

\emailAdd{mohamed.younes.sassi@desy.de}
\emailAdd{gudrid.moortgat-pick@desy.de}

\abstract{The Next-to-Two-Higgs-Doublet model (N2HDM) has a rich vacuum structure where multiple electroweak (EW) breaking minima, as well as CP and electric-charge breaking minima, can coexist. These minima can be deeper than the electroweak vacuum  $v_{ew} \approx 246\text{ GeV}$ of our universe, making our vacuum metastable. In such a case, one needs to calculate the tunneling rate from the EW vacuum to the deeper minimum. If the lifetime of the EW vacuum is longer than the universe's age, our vacuum is deemed long-lived, and the parameter point is, in principle, allowed. If the decay rate is smaller than the universe's age, then our vacuum is unstable and the parameter point is ruled out. However, domain walls (DW) in the N2HDM can substantially alter this picture. We show in this work that inside the DW, the barrier between our electroweak minimum and the deeper minimum can disappear, leading the scalar fields to classically roll over to the deeper minimum that nucleates inside the DW and then expands outside of it everywhere in the universe. We show that such behavior can happen to parameter points where the lifetime of our minimum is even several orders of magnitude larger than the age of the universe. Such parameter points with a metastable EW minimum are ruled out.}

\begin{document} 

\maketitle
\flushbottom

\section{Introduction}

In the Standard Model (SM) of particle physics, quarks, leptons, gauge, and scalar bosons acquire a mass via the Higgs mechanism, relying on the scalar field of the Higgs boson. After electroweak symmetry breaking in the early universe, the Higgs field acquires a vacuum expectation value (VEV) corresponding to a minimum of the scalar potential. In the SM, this minimum corresponds to a VEV $v_{sm} \approx 246 \text{ GeV}$. 
At tree-level, the SM electroweak (EW) minimum is stable. However, when higher-order corrections are included in the parameters of the SM Higgs potential, the potential can develop another deeper minimum at high scales, rendering the SM EW minimum to be a local minimum and thus metastable \cite{Hiller:2024zjp, Bednyakov:2015sca}. Current experimental results for the top quark mass, the strong coupling constant, and the W boson mass favor this scenario \cite{CMS:2019esx}, and the electroweak (EW) vacuum in the SM could, therefore, be metastable but long-lived. 

When considering an extended scalar sector, the addition of extra degrees of freedom can lead, already at tree-level, to the existence of several minima of the scalar potential other than the EW minimum \cite{Ferreira:2019iqb, Lahiri:2024rya, Barroso:2005sm, Barroso:2012mj, Barroso:2013awa, Ivanov:2006yq, Branchina:2018qlf, Ivanov:2007de}. The EW minimum is then stable if it is the global minimum of the potential, metastable when the tunneling rate between the EW minimum and deeper minima of the potential leads to an EW minimum with a lifetime larger than the age of the universe, and unstable when the lifetime of the EW minimum is smaller than the age of the universe. Obviously, an unstable EW minimum is ruled out, and the vacuum (meta)stability of the EW vacuum can be used as a strong constraint for the viability of models with an extended scalar sector \cite{Ferreira:2019iqb, Lahiri:2024rya, Barroso:2005sm, Barroso:2012mj, Barroso:2013awa, Ivanov:2006yq, Branchina:2018qlf, Ivanov:2007de}. 

In this work, we consider the Next-to-Two-Higgs-Doublet-Model (N2HDM) where the SM Higgs scalar doublet is extended by an extra scalar doublet charged under $\text{SU(2)}_L\times\text{U(1)}_Y$, and by a real singlet scalar. The N2HDM is a well-motivated extension of the scalar sector since it can help to overcome several shortcomings in the SM. This includes the possibility of EW first-order phase transitions needed for EW baryogenesis, which provides a mechanism to generate the observed matter-antimatter asymmetry in the early universe \cite{Biekotter:2021ysx}. One can also use the N2HDM to generate several dark matter candidates \cite{Glaus:2022rdc, Engeln:2020fld}. The N2HDM also provides a useful benchmark model for experimental searches for new scalar bosons at colliders and can accommodate the observed 95 GeV excess in CMS and ATLAS \cite{Biekotter:2019kde, Biekotter:2022jyr, Heinemeyer:2021msz, ATLAS:2018xad, ATLAS:2024itc, CMS:2018cyk, CMS:2024yhz, Janot:2024ryq, LEPWorkingGroupforHiggsbosonsearches:2003ing}.

Since models with multiple Higgs doublets can induce experimentally highly constrained flavor-changing neutral currents at tree-level \cite{Branco:2011iw}, one usually introduces a $Z_2$ discrete symmetry to the model in order to force each Higgs doublet to only couple to a fermion generation at a time and therefore avoid this constraint. An interesting consequence of the spontaneous breaking of a discrete symmetry in the early universe is that it leads to the formation of topological defects known as domain walls (DW). These defects are stable field configurations that interpolate between two degenerate minima belonging to different disconnected sectors of the vacuum manifold of the model \footnote{These minima are degenerate and lead to the same vacuum of the theory. No stable domain wall field configurations can be formed between e.g. the EW minimum and deeper minima of the potential.}, which are related by the discrete symmetry. The formation of domain walls in the early universe is problematic from a cosmological perspective, since they quickly dominate the energy of the universe \cite{Kibble:1976sj, Zeldovich:1974uw, VILENKIN1985263}. One way to avoid this so-called domain wall problem is to add terms in the potential that explicitly break the discrete symmetry \cite{vachaspati_2023, VILENKIN1985263,Dvali:1995cc}. This then breaks the degeneracy between both minima related by the discrete symmetry. The regions of the universe populated by the false vacuum will shrink until they disappear, leading to the annihilation of the domain wall network. Although this problem has led in the past to a lack of interest in domain walls compared to other cosmologically benign topological defects such as cosmic strings, the topic of domain walls has experienced a renewed interest in the last few years since one can use an annihilating domain wall network as a source for gravitational waves which could fit well the data from Pulsar Timing Arrays (PTA) observations \cite{NANOGrav:2023gor, Gouttenoire:2023ftk}. One can also embed domain walls in the context of electroweak baryogenesis to generate the observed asymmetry between matter and antimatter in the universe \cite{Brandenberger:1994mq, Sassi:2024cyb, Schroder:2024gsi, Azzola:2024pzq, Vanvlasselaer:2024vmi}, as well as inducing the electroweak symmetry breaking in the early universe since the walls can act as impurities that catalyze the nucleation of electroweak vacuum bubbles \cite{Blasi:2022woz,Agrawal:2023cgp}. In our work, we consider the formation of domain walls related to the breaking of the discrete symmetry $Z'_2$ that acts on the real singlet scalar of the N2HDM. We assume that this symmetry is only slightly broken (for example, by quantum gravity effects \cite{Fichet:2019ugl}). This avoids the domain wall problem, while still allowing the use of the $Z'_2$ invariant N2HDM phenomenology where the discrete symmetry is exact.

The vacuum stability of the $Z'_2$ invariant N2HDM was extensively investigated \cite{Ferreira:2019iqb, Lahiri:2024rya}. It was found that large regions of the viable parameter space in the N2HDM can be metastable, with the EW vacuum being very long-lived \cite{Ferreira:2019iqb, Lahiri:2024rya}. In such a case, the potential barrier between the EW minimum and the global minimum is very large, leading to an extremely small tunneling rate between the EW vacuum and the global one. However, these studies didn't take into account the possible presence of domain walls in the N2HDM. Due to the coupling of the singlet and doublet scalar fields, the potential for the scalar doublets in the background of the singlet domain wall will be dependent on the position with respect to the core of the domain wall. Inside the core of the defect, the VEV of the singlet field vanishes, and the effective quadratic terms for the doublet fields will be different. Because of this, the potential barrier to the global minimum of the potential can be different or even vanish, leading to the nucleation of the global minimum inside the wall. Due to the difference in the potential energy, the global minimum will then expand outside of the wall, and the universe is eventually populated with the global minimum instead of the EW minimum. Since the VEV of this global vacuum is generally different from the EW minimum, the masses of all elementary particles would be different, which leads to such parameter points being ruled out. 

Our work is organized as follows: in section \ref{section2}, we introduce the N2HDM potential and the used notation as well as the different types of minima that can occur in the model. In section \ref{section3}, we briefly discuss the vacuum stability of the N2HDM, before we study in section \ref{section4} how domain walls can lead to the decay of very long-lived EW minima via a classical rollover triggered inside the core of the wall. In section \ref{section5}, we describe some phenomenological scenarios to demonstrate the possibility of using the mechanism of vacuum decay via domain walls and rule out large regions of the otherwise viable parameter points in the N2HDM. We summarize and conclude our work in section \ref{section6}. 

\section{The Next-to-Two-Higgs-Doublet Model}\label{section2}
This section briefly introduces the Next-to-Two-Higgs-Doublet model and the needed notation used in our work. For a comprehensive review of this model, the reader is referred to \cite{Muhlleitner:2016mzt, Muhlleitner:2017dkd,Chen:2013jvg}, which includes the phenomenology of this model, and \cite{Biekotter:2021ysx} for a discussion of its thermal history. \\
In the N2HDM, the standard model Higgs sector is extended with an extra $SU(2)_L \times U(1)_Y$ doublet $\Phi_2$ and an additional real singlet $\Phi_s$. The scalar sector potential is given by:
\begin{align}
   \notag V_{N2HDM} &= m^2_{11}\Phi^{\dagger}_1\Phi_1 + m^2_{22}\Phi^{\dagger}_2\Phi_2 -  m^2_{12}(\Phi^{\dagger}_1\Phi_2 + h.c.) + \frac{\lambda_1}{2}\bigl(\Phi^{\dagger}_1\Phi_1\bigr)^2 + \frac{\lambda_2}{2}\bigl(\Phi^{\dagger}_2\Phi_2\bigr)^2 \\ \notag 
    &   + \lambda_3\bigl(\Phi^{\dagger}_1\Phi_1\bigr)\bigl(\Phi^{\dagger}_2\Phi_2\bigr)  + \lambda_4\bigl(\Phi_1^{\dagger} \Phi_2\bigr)\bigl(\Phi_2^{\dagger} \Phi_1\bigr) 
    +\biggl[\frac{\lambda_5}{2}\bigl(\Phi_1^{\dagger} \Phi_2\bigr)^2 + h.c\biggr] \\ \notag
    & + \frac{m^2_S}{2}\Phi^2_s + \frac{\lambda_6}{8}\Phi^4_s + \frac{\lambda_7}{2}\Phi^2_s(\Phi^{\dagger}_1\Phi_1) + \frac{\lambda_8}{2}\Phi^2_s(\Phi^{\dagger}_2\Phi_2) \\ & + \biggl[a_1\Phi_s + a_3 \Phi^3_s + b_1(\Phi^{\dagger}_1\Phi_1)\Phi_s + b_2(\Phi^{\dagger}_2\Phi_2)\Phi_s + c_1 (\Phi^{\dagger}_1\Phi_2\Phi_s + h.c.) \biggr] .
\label{eq:treepot}    
\end{align}
In order to avoid flavor-changing neutral currents \cite{Branco:2011iw}, one imposes a $Z_2$ symmetry that acts on the scalar fields in the following way:
\begin{align}
    \Phi_1 \rightarrow \Phi_1, && \Phi_2 \rightarrow -\Phi_2, && \Phi_s \rightarrow \Phi_s.
\end{align}
When this symmetry is extended to the Yukawa sector, the model can be classified into different types depending on how the fermions transform under $Z_2$ (see Tab \ref{Tab:2hdmtypes}).
\begin{table}[t]
\centering
\begin{tabular}{ |c||c|c|c||c|c|c|c|c|  }
 \hline
 & u-type & d-type & leptons & Q & $u_R$ & $d_R$ & L & $l_R$\\
 \hline \hline
Type 1  & $\Phi_2$  & $\Phi_2$ & $\Phi_2$ & +  & - & - & + & - \\
Type 2 & $\Phi_2$ & $\Phi_1$  & $\Phi_1$ & + & - & + & + & + \\
Type 3 (lepton specific)  & $\Phi_2$  & $\Phi_2$ & $\Phi_1$ & +  & - & - & + & + \\
Type 4 (Flipped)  & $\Phi_2$   & $\Phi_1$ & $\Phi_2$ & +  & - & + & + & - \\
\hline
\end{tabular}
\caption{Types of Yukawa couplings between the fermions and the scalars in the N2HDM and the charges of the fermions under the $Z_2$ symmetry \cite{Branco:2011iw}. $Q$ and $L$ denote left-handed quark and lepton $SU(2)_L$ doublets while $u_R$, $d_R$ and $l_R$ denote $SU(2)_L$ right-handed singlets.}
\label{Tab:2hdmtypes}
\end{table}
This symmetry is softly broken by the terms $m^2_{12}(\Phi^{\dagger}_1\Phi_2 + h.c)$ and $c_1 (\Phi^{\dagger}_1\Phi_2\Phi_s + h.c.)$. When the parameters $a_1$, $a_3$, $b_1$, $b_2$, and $c_1$ are zero, the potential also allows for a discrete symmetry $Z'_2$, which only acts on the singlet:
\begin{equation}
    \Phi_s \rightarrow -\Phi_s.
\end{equation}
In this work, we aim to study the influence of DW field configurations on EW vacuum decay to deeper minima of the potential in the case of the standard N2HDM with an exact $Z'_2$ symmetry\footnote{In the case of a vanishing vacuum expectation value of the real singlet scalar, this symmetry can be used to forbid the decay of the real singlet scalar and make it a viable dark matter candidate \cite{Glaus:2022rdc, Engeln:2020fld}.}. We therefore assume that the $Z'_2$ breaking terms avoid the cosmological domain wall problem, but are still very small and therefore can be neglected in the discussion of the vacuum structure and the phenomenology of the model. This can occur, for example, if the relevant scale of $Z'_2$ breaking is the Planck scale since quantum gravity effects are known to break any discrete symmetries \cite{Gouttenoire:2025ofv}. In such a case, one can avoid the cosmological domain wall problem while still having a similar phenomenology to the standard N2HDM potential (where $a_1$, $a_3$, $b_1$, $b_2$, and $c_1$ are set to zero). 

After electroweak and $Z'_2$ symmetries breaking, the scalar doublets and singlet acquire a vacuum expectation value. The most general vacuum configuration can be written as:
\begin{align}
   \langle \Phi_1 \rangle = \text{U} \langle \tilde{\Phi}_1 \rangle = \text{U} \dfrac{1}{\sqrt{2}}
    \begin{pmatrix}
          0 \\      v_1
     \end{pmatrix},      
&& \langle \Phi_2 \rangle = \text{U} \langle \tilde{\Phi}_2 \rangle = \text{U} \dfrac{1}{\sqrt{2}}
      \begin{pmatrix}
     v_+ \\
     v_2e^{i\xi}
      \end{pmatrix} , && \langle \Phi_s \rangle = v_s,
\label{eq:vacuumform}      
\end{align}
where U is an element of the $\text{SU(2)}_L\times\text{U(1)}_Y$ group that is given by:
\begin{equation}
        \text{U} = e^{i\theta} \text{exp}\biggl(i\dfrac{\tilde{g}_i\sigma_i}{2v_{sm}}\biggl),
\label{eq:EWmatrix}        
\end{equation}
with $\theta$ and $\tilde{g}_i$ denoting the Goldstone modes of the scalar doublets, $\sigma_i$ the Pauli matrices and $v_{sm} \approx 246 \text{ GeV} $ the standard model vacuum expectation value.

The scalar doublets admit three possible types of vacua, which we discuss extensively in the next chapter. The most general one, where $v_+ \neq 0$, breaks the electromagnetism symmetry $U(1)_{em}$ and gives a mass to the photon. Consequently, such vacua are physically not allowed at present time. The second type occurs when the phase between the two scalar doublets $\xi$ does not vanish. Such a vacuum is CP-violating since it generates an imaginary mass to the fermions via the Yukawa sector. Due to constraints from electric dipole moment experiments \cite{Roussy:2022cmp, Chupp:2017rkp}, such CP-violating vacua should have very small values for $\xi$ to be realized in nature. The third type is the neutral vacuum, occurring when $v_+ = 0$ and $\xi = 0$. Concerning the real scalar singlet, we consider in this study the case where the singlet scalar acquires a vacuum expectation value $v_s \neq 0$, which breaks $Z'_2$ spontaneously and leads to the formation of domain walls in the early universe. 

After electroweak symmetry breaking (EWSB), the particle spectrum of the N2HDM includes 3 CP-even Higgs particles with masses denoted as $m_{h_1}$, $m_{h_2}$ and $m_{h_3}$, one CP-odd particle with mass $m_A$ and two charged Higgs bosons $m_{H^{\pm}}$. It is more advantageous to express the potential parameters in terms of physical quantities such as the masses of the physical particles and $\tan(\beta) = v_2/v_1$. This is achieved by diagonalizing the mass matrix $M^2_\rho$ (see (\ref{eq:massmatrix})) given in the interaction basis $(\rho_1, \rho_2, \rho_3)$, where $\rho_{1,2,3}$ correspond to field expansions around the neutral vacua $v_{1,2,s}$ in (\ref{eq:vacuumform}):
\begin{equation}
M_{\rho}^{2}=\left(\begin{array}{ccc}
v^{2}\lambda_{1}\cos(\beta)^{2}+m_{12}^{2}\, tan(\beta) \,\, &
v^{2}\lambda_{345}\,\cos(\beta)\,\sin(\beta)-m_{12}^{2}\,\, &
v\,v_{s}\lambda_{7}\,\cos(\beta)\\
v^{2}\lambda_{345}\,\cos(\beta)\,\sin(\beta)-m_{12}^{2} \,\, &
v^{2}\lambda_{2}\,\sin(\beta)^{2}+m_{12}^{2}/tan(\beta)\,\,&
v\,v_{s}\lambda_{8}\,\sin(\beta)\\
v\,v_{s}\lambda_{7}\,\cos(\beta)\,\, & v\,v_{s}\lambda_{8}\,\sin(\beta)\, \,&
v_{s}^{2}\,\lambda_{6}
\end{array}\right),
\label{eq:massmatrix}
\end{equation}
where $v^2 = v^2_1 + v^2_2$. This mass matrix is diagonalized using a rotation matrix R which fulfills the requirement $RM^2_{\rho}R^{T} = diag(m^2_{h_1}, m^2_{h_2}, m^2_{h_3})$, where the masses $m_{h_{1,2,3}}$ correspond to the masses of the CP-even Higgs bosons in the physical mass basis $(h_1, h_2, h_3)$. The diagonalizing matrix R is parametrized using the mixing angles $\alpha_1$, $\alpha_2$ and $\alpha_3$ as:
\begin{equation}
R = 
\begin{pmatrix}
c(\alpha_1)c(\alpha_2)  & s(\alpha_1)c(\alpha_2) & s(\alpha_2)\\
-\bigl(c(\alpha_1)s(\alpha_2)s(\alpha_3) + s(\alpha_1)c(\alpha_3)\bigr) & c(\alpha_1)c(\alpha_3) - s(\alpha_1)s(\alpha_2)s(\alpha_3) & c(\alpha_2)s(\alpha_3) \\
-c(\alpha_1)s(\alpha_2)c(\alpha_3) + s(\alpha_1)s(\alpha_3) & -\bigl(c(\alpha_1)s(\alpha_3) + s(\alpha_1)s(\alpha_2)c(\alpha_3) \bigr) & c(\alpha_2)c(\alpha_3)
\end{pmatrix},
\label{eq:Rmatrix}
\end{equation}
where $c(\alpha_i)$ denotes $\cos(\alpha_i)$ and $s(\alpha_i)$ denotes $\sin(\alpha_i)$. The values of the mixing angles are constrained between $-\pi/2$ and $\pi/2$. We adopt the conventional mass hierarchy $m_{h_1} < m_{h_2} < m_{h_3}$. Note that the interaction basis $(\rho_1, \rho_2, \rho_3)$ is related to the physical mass basis $(h_1, h_2, h_3)$ by:
\begin{equation}
    \begin{pmatrix} h_1 \\ h_2  \\  h_3  \end{pmatrix} = R \begin{pmatrix} \rho_1 \\ \rho_2  \\  \rho_3  \end{pmatrix}.
\end{equation}
One can then relate the potential parameters and the masses of the scalars in the N2HDM using the following formulas: 
\begin{align}
\lambda_1 &= \frac{1}{v^2_1}\biggl(-m^2_{12}\tan(\beta) + \sum_i m^2_{h_i}R^2_{i1}\biggr), \\
\lambda_2 &= \frac{1}{v^2_2}\biggl(-\frac{m^2_{12}}{\tan(\beta)} + \sum_i m^2_{h_i}R^2_{i2}\biggr), \\
\lambda_3 &= \frac{1}{v_1v_2}\biggl(m^2_{12} + \sum_i R_{i2}R_{i1}m^2_{h_i}\biggr) - \lambda_4 - \lambda_5, \\
\lambda_4 &= \frac{m_{12}}{v_1v_2} - 2 \frac{m^2_{H^{\pm}}}{v^2} + \frac{m^2_{A}}{v^2}, \\
\lambda_5 &= \frac{m^2_{12}}{v_1v_2} - \frac{m^2_A}{v^2}, \\
\lambda_6 &= \frac{1}{v^2_s}\biggl( R^2_{i3}m^2_{H_i}\biggr), \\
\lambda_7 &= \frac{1}{v_1v_s}\biggl( R_{i3}R_{i1}m^2_{h_i} \biggr), \label{eq:lambda7} \\
\lambda_8 &= \frac{1}{v_2v_s}\biggl( R_{i3}R_{i2}m^2_{h_i} \biggr), \label{eq:lambda8} \\
m^2_{11} &= m^2_{12}\tan(\beta) - \frac{\lambda_1}{2}v^2_1 - \biggl(\frac{\lambda_3 + \lambda_4 + \lambda_5}{2}\biggr)v^2_2 - \frac{\lambda_7}{2}v^2_s, \\
m^2_{22} &= \frac{m^2_{12}}{\tan(\beta)} - \frac{\lambda_2}{2}v^2_2 - \biggl(\frac{\lambda_3 + \lambda_4 + \lambda_5}{2}\biggr)v^2_1 - \frac{\lambda_8}{2}v^2_s, \\
m^2_s &= -\frac{\lambda_6}{2}v^2_s - \frac{\lambda_7}{2}v^2_1 - \frac{\lambda_8}{2}v^2_2.
\end{align}
In the next chapter, we briefly discuss the vacuum structure of the N2HDM and the different possibilities to obtain minima that lie deeper in the potential compared to the physical EW minimum.

\section{Vacuum Instabilities in the N2HDM}\label{section3}

In this section, we discuss the concept of vacuum instability and summarize the results obtained in \cite{Ferreira:2019iqb}, using the potential of the $Z'_2$ symmetric N2HDM in (\ref{eq:treepot}).   

In our work we consider parameter points where the physical vacuum has an EW vacuum for the Higgs doublets $v_{ew} = \sqrt{v^2_1+ v^2_2} \approx 246 \text{ GeV}$ leading to the observed masses of the weak gauge fields as well as a non-zero VEV for the singlet scalar $v_s \neq 0$. We follow the notation in \cite{Ferreira:2019iqb} and denote this minimum as $\mathcal{N}s$ :
\begin{align}
   \langle \Phi_1 \rangle_{\mathcal{N}s} = \dfrac{1}{\sqrt{2}}
    \begin{pmatrix}
          0 \\      v_1
     \end{pmatrix},      
&& \langle \Phi_2 \rangle_{\mathcal{N}s} =  \dfrac{1}{\sqrt{2}}
      \begin{pmatrix}
     0 \\
     v_2
      \end{pmatrix} , && \langle \Phi_s \rangle_{\mathcal{N}s} = v_s.
\label{eq:ewvacuum}      
\end{align}
It was found in \cite{Ferreira:2019iqb} that for several parameter points of the model, the vacuum $\mathcal{N}s$ is not the global minimum, and different types of vacua can be deeper. 
This includes: 

\noindent
\textbullet\ Electric-charge breaking vacua $\mathcal{CB}$:
\begin{align}
   \langle \Phi_1 \rangle_\mathcal{CB} = \dfrac{1}{\sqrt{2}}
    \begin{pmatrix}
          0 \\      c_1
     \end{pmatrix},      
&& \langle \Phi_2 \rangle_\mathcal{CB} =  \dfrac{1}{\sqrt{2}}
      \begin{pmatrix}
     c_+ \\
     c_2
      \end{pmatrix} , && \langle \Phi_s \rangle_\mathcal{CB} = 0.
\label{eq:cbvacuum}      
\end{align}
\textbullet\ CP breaking vacua $\mathcal{CP}$:
\begin{align}
   \langle \Phi_1 \rangle_\mathcal{CP} = \dfrac{1}{\sqrt{2}}
    \begin{pmatrix}
          0 \\      \bar{v}_1
     \end{pmatrix},      
&& \langle \Phi_2 \rangle_\mathcal{CP} =  \dfrac{1}{\sqrt{2}}
      \begin{pmatrix}
     0 \\
     \bar{v}_2e^{i\xi}
      \end{pmatrix} , && \langle \Phi_s \rangle_\mathcal{CP} = 0.
\label{eq:deeperewvacuum}      
\end{align}
\textbullet\ Deeper neutral EW breaking vacua leading to different gauge boson masses $v'_{ew} = \sqrt{v'^2_1+ v'^2_2} \neq 246 \text{ GeV}$ and a different $v_s$ (denoted as $\mathcal{N}'s$) or a vanishing $v_s = 0$ (denoted as $\mathcal{N'}$):
\begin{align}
   \langle \Phi_1 \rangle_{\mathcal{N}'} = \dfrac{1}{\sqrt{2}}
    \begin{pmatrix}
          0 \\      v'_1
     \end{pmatrix},      
&& \langle \Phi_2 \rangle_{\mathcal{N}'} =  \dfrac{1}{\sqrt{2}}
      \begin{pmatrix}
     0 \\
     v'_2
      \end{pmatrix} , && \langle \Phi_s \rangle_{\mathcal{N}'} = 0, && \langle \Phi_s \rangle_{\mathcal{N}'s} = v'_s.
\label{eq:cpvacuum}      
\end{align}
Electric charge and CP-breaking vacua with a non-zero VEV for $v_s$ cannot be deeper than the $\mathcal{N}s$ (see \cite{Ferreira:2019iqb} for a detailed discussion), and we therefore don't consider them in this work.

The existence of minima deeper than our EW minimum requires a careful study of the metastability of the vacuum. This requires the calculation of the tunneling rate from the EW vacuum to the deeper vacuum. This rate per unit volume is related to the bounce action B by \cite{Hollik:2018wrr}:
\begin{equation}
    \Gamma = Ke^{-B},
\label{eq:tunnelingrate}    
\end{equation}
where K is a dimensionful parameter that has subdominant effects on the value of the tunneling rate. The bounce action B for a scalar field configuration $\phi$ is obtained by finding the stationary point of the Euclidean action \cite{Hollik:2018wrr}:
\begin{equation}
    B = 2\pi^2 \int^{\infty}_0 \rho^3 \dv \rho \biggl[ \dfrac{1}{2} \biggl( \dv \rho \phi_B(\rho) \biggr)^2 + V(\phi_B(\rho)) \biggr],
\end{equation}
where $\rho=\sqrt{t^2-x^2 - y^2 - z^2}$ denotes the spacetime variable, V is the scalar field potential, and $\phi_B(\rho)$ is the field bounce solution of the Euclidean equation of motion:
\begin{equation}
    \dv[2] {\phi} {\rho} + \dfrac{3}{\rho}\dv {\phi}{\rho} = \pdv{V}{\phi},
\end{equation}
which is solved using the boundary conditions:
\begin{equation}
    \phi(\infty) = \phi_v \quad\mathrm{and}\quad   \biggl(\dv{\phi}{\rho}\biggr)_{\rho=0}= 0,
\end{equation}
with $\phi_v$ denoting the metastable vacuum. Calculating this bounce action and determining the tunneling rate is usually done using numerical tools such as e.g. \texttt{EVADE} \cite{Hollik:2018wrr}, which we use throughout our work. Since the tunneling rate (\ref{eq:tunnelingrate}) to the global minimum is exponentially suppressed with the value of $B$, the bounce action $B$ predominantly determines the fate of the EW vacuum. By comparing the age of the universe to the value of $\Gamma$, one determines the intervals where the EW vacuum is unstable or metastable \cite{Hollik:2018wrr}. For parameter points with $B>440$, the transition from the EW vacuum to the deeper vacuum takes longer than the age of the universe \cite{Hollik:2018wrr}. Therefore, our EW vacuum is then deemed metastable and long-lived, and the parameter point is not ruled out. In case $390<B<440$, the fate of the EW vacuum is uncertain, and for $B<390$, the EW vacuum is short-lived and unstable, rendering such parameter points unphysical \cite{Hollik:2018wrr}. 

This analysis of vacuum stability is valid for the regions of the universe that are far from the domain wall of the singlet scalar. The dangerous types of minima ($\mathcal{CB}$, $\mathcal{CP}$ and $\mathcal{N'}$ defined in (\ref{eq:cbvacuum}), (\ref{eq:cpvacuum}), and (\ref{eq:deeperewvacuum})), which lie deeper in the potential than the EW minimum, have a vanishing $v_s$. Since the singlet scalar field inside the core of the domain wall has to vanish, it is crucial to check whether the domain walls can facilitate the decay of the EW vacuum to those deeper vacua, rendering those metastable parameter points unphysical.

\begin{table}
  \centering
  \begin{tabular}{lcccccc}
        & $m_{H_y},m_{H_z},m_A$ & $m_{H^\pm}$ & $\tan\beta$ & $m_{12}^2$            & $v_S$ \\
     \hline
      \hline
    min & 30 GeV                & 150 GeV     & 0.8         & 0 GeV$^2$             & 1 GeV \\
    max & 1.5 TeV               & 1.5 TeV     & 20          & $5\times10^5$ GeV$^2$ & 3 TeV \\
     \hline
     \hline
  \end{tabular}
  \caption{Input parameters used for the scan with \texttt{ScannerS}. The mixing angles $\alpha_{1,2,3}$ were not constrained. The mass $m_{H_x} = 125.09\text{ GeV}$.}\label{tab:scanold}
\end{table}

Before discussing the role of domain walls in inducing the decay of long-lived EW vacua to deeper vacua, we briefly summarize some phenomenological aspects of the vacuum stability and instability and their influence on the allowed parameter space of the N2HDM. This is done following the numerical approach used in \cite{Ferreira:2019iqb}, and we refer the reader to that work for a comprehensive discussion. We perform a parameter scan of $10^5$ parameter points of the N2HDM of type 1 (see Tab \ref{Tab:2hdmtypes}), which fulfill both theoretical constraints of perturbative unitarity and boundedness from below, as well as the experimental constraints from precision electroweak variables, collider constraints from Higgs searches and Higgs measurements, as well as flavor constraints (see section \ref{appendixconstraints}). We also impose the condition of electroweak and $Z'_2$ symmetry restoration at high temperatures (see section \ref{appendixconstraints}) in order to ensure the formation of the singlet domain walls in the early universe. All these constraints were implemented in \texttt{ScannerS} \cite{Muhlleitner:2020wwk}, which generates random parameter points satisfying all these conditions. We follow the analysis in \cite{Ferreira:2019iqb} and take the same range of model variables for the generated parameter points (see Tab \ref{tab:scanold}). For each parameter point, the stability of the EW vacuum is verified against the existence of deeper vacua using \texttt{EVADE}. We find that for this random parameter scan, $33\%$ of the viable parameters exhibit a deeper vacuum than the EW one. These vacua can be of the $\mathcal{N}'$, $\mathcal{CB}$ and $\mathcal{CP}$ types. Almost $25\%$ of the viable parameter points have a long-lived EW vacuum that has a lifetime larger than the age of the universe, and $8\%$ of the parameter points exhibit a short-lived EW vacuum that decays to the deeper vacuum and therefore is a dangerous minimum. The deeper vacua that were found in this scan had a vanishing $v_s$ and therefore the rate of decay of the EW minimum will be enhanced in the vicinity and inside the singlet wall (where $v_s(0) = 0$), as will be shown later. 

\begin{figure}[h]
     \centering
     \subfloat[]{\includegraphics[width=0.5\columnwidth]{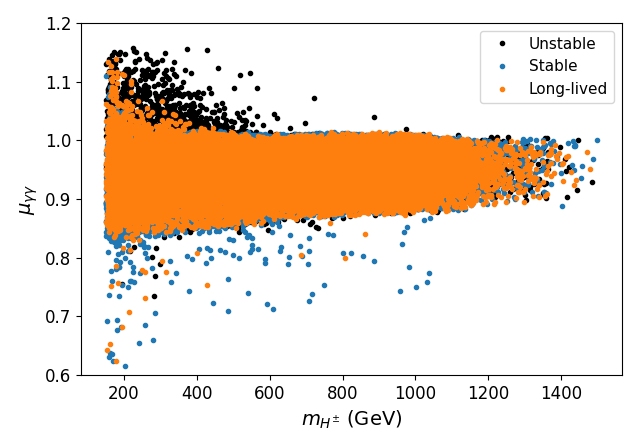}
     \label{subfig:gammamumu}}
     \subfloat[]{\includegraphics[width=0.5\columnwidth]{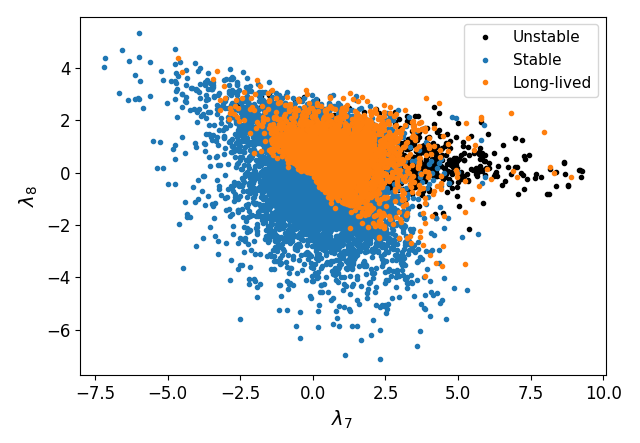}
        \label{subfig:l7l8}}
\caption{EW vacuum (in)stability for the random parameter scan. (a) The stability of the EW vacuum as a function of the signal ratio $\mu_{\gamma\gamma}$ and the charged Higgs mass $m_H^{\pm}$. (b) The stability of the vacuum as a function of the potential variables $\lambda_7$ and $\lambda_8$.} 
\label{fig:vacinstability}
\end{figure}

The presence of deeper minima than the EW one in this scan is random and does not exhibit strong correlations with the variables of the model. This was also found in \cite{Ferreira:2019iqb}, where it was shown that it is possible to obtain some constraints on the signal strength $\mu_{\gamma\gamma}$ defined as:
\begin{equation}
    \mu_{\gamma\gamma} = \dfrac{\sigma(pp \rightarrow h_{125})\text{BR}(h_{125} \rightarrow \gamma\gamma)}{\sigma(pp \rightarrow h_{sm})\text{BR}(h_{sm} \rightarrow \gamma\gamma)},
\end{equation}
which describes the decay rate of the $h_{125}$ mass eigenstate to a pair of photons normalized to the rate of the SM Higgs decay to two photons. The results are shown in Figure \ref{fig:vacinstability} where we differentiate between parameter points having a stable EW vacuum (blue), parameter points having a long-lived EW vacuum (orange), and parameter points having a short-lived unstable EW vacuum (black). The points with an unstable EW vacuum are plotted first (black), then stable points (blue), and finally metastable ones on top (orange).

In the case of $\mu_{\gamma\gamma}$ dependence (shown in Figure \ref{subfig:gammamumu}), we find that the regions of vacuum stability, metastability, and instability mostly overlap. For small masses $m_{H^\pm}$ between $300 \text{ GeV}$ and $600 \text{ GeV} $ and $\mu_{\gamma\gamma}>1$, we find that the EW vacuum is unstable and therefore one can exclude that region of the parameter space on vacuum stability arguments. We also observe (see Figure \ref{subfig:l7l8}) that the EW vacuum for negative $\lambda_7$ and $\lambda_8$ is always stable, and the EW minimum is then the global minimum. The absence of a deeper minimum in this case can be explained by the fact that for $\lambda_{7,8}<0$, the effective mass terms of the Higgs doublets receive a large positive contribution in the direction where $\phi_s \rightarrow 0$, due to $\lambda_7 \abs{\Phi_1}^2 \phi^2_s + \lambda_8 \abs{\Phi_2}^2 \phi^2_s$ in the scalar potential (\ref{eq:treepot}) vanishing. In this case, the potential of the Higgs doublets approaches the local extremum $(v_1,v_2) = (0,0)$, which always lies higher in the potential than the EW minimum (at zero temperature).

Even though there is no strong correlation between short-lived EW vacua and model parameters, vacuum instability can be used to exclude a large number of parameter points of the model, which would otherwise be viable from a theoretical and experimental point of view. We also find that a large set of the parameter points of the random scan shows the presence of long-lived EW vacua. In principle, these metastable vacua are physically allowed since their lifetime is (much) larger than the age of the universe. However, we found that the deeper vacua of these parameter points have a vanishing $v_s = 0$. In the case of a homogeneous vacuum everywhere in the universe where every point in space, after EWSB, falls into the EW vacuum $\mathcal{N}s$, the metastability of the EW vacuum is ensured. However, the presence of domain walls makes the singlet VEV $v_s(x)$ space-dependent, with regions of the universe inside the domain wall having $v_s(0)=0$. It is therefore important to check whether the presence of the walls will induce the decay into the deeper vacuum, making the long-lived EW vacuum unstable. In the next chapter, we solve the domain wall solutions in the N2HDM and check the fate of long-lived EW vacua inside and in the vicinity of the domain wall.

\section{Domain Walls in the N2HDM inducing EW vacuum decay}\label{section4}
The spontaneous breaking of the $Z'_2$ symmetry leads to the formation of cosmic domain walls in the early universe \cite{Kibble:1976sj}. Due to the $Z'_2$ symmetry, the two vacua $v_s$ and $-v_s$ are degenerate\footnote{As mentioned before, we assume the $Z'_2$ soft breaking terms in the Lagrangian to be very small, making both vacua effectively degenerate.} and therefore, have the same probability of occurring in the early universe. Different patches of the universe would then acquire different signs for $v_s$, with $v_s=0$ inside the walls separating these domains. In order to obtain the field profile of the domain walls, one solves the equation of motion of the scalar fields $\phi_i$ with $i={1,2,+,s}$ (as defined in (\ref{eq:vacuumform})), which will minimize the energy density of the vacuum configuration:
\begin{equation}
    \pdv[2]{\phi_i}{t} - \pdv[2]{\phi_i}{x} + d\pdv{\phi_i}{t}  + \pdv{V_{N2HDM}}{\phi_i} = 0,
\label{eq:eom}    
\end{equation}
where d is a friction term needed in order to relax the initial energy of the field configuration into the lowest energy solution. This friction term is related to both the expansion rate in the early universe and the friction induced by the interaction of the domain wall with the primordial thermal plasma. In our work, we do not explicitly estimate the value of $d$, but we use it in order to relax the field configuration into its lowest energy configuration. For $d=0$, we would obtain large gradients in the fields configuration which disturb the numerical solution of this system of equations\footnote{The value of $d$ can be relevant in the case when the 2HDM potential inside the wall ($\phi_s=0$) has multiple minima, as we will discuss later.}.
The boundary conditions are taken to be $\phi_s = -v_s$ at $x=-\infty$ and $\phi_s = v_s$ at $x=+\infty$ for the singlet field and $\phi_{1,2} = v_{1,2}$ at $\pm \infty$ where $v_{1,2}$ denote the EW vacuum ($\sqrt{v^2_1 + v^2_2}\approx 246 \text{ GeV}$). The initial field profiles are taken to be a $\phi_s(x) = v_s \tanh(\delta x)$ for the singlet field, where $\delta$ denotes the inverse width of the DW, and a constant profile $\phi_{1,2}(x) = v_{1,2}$ for the doublets. We obtain the static field configuration by solving the system of differential equations numerically using the Euler method with spatial derivatives calculated via the three-point central difference formula. 

We solve the coupled system of differential equations for the potential at temperature $T=0$. This is done for two reasons. In this work, we only consider the possibility of EW vacuum decay induced by the presence of domain walls in order to constrain the parameter points of the N2HDM. Therefore, we don't need to know exactly at which temperature, after EW symmetry breaking, the decay to the deeper vacuum occurs. Since this decay will necessarily rule out such parameter points, determining the precise thermal evolution of such parameter points is not phenomenologically relevant for this study. Performing large parameter scans and determining a precise thermal evolution of the finite-temperature potential of the N2HDM using available dedicated tools such as BSMPT \cite{Basler:2024aaf} is a computationally time-consuming task. For simplicity and speed of the calculation, we use the potential at $T=0$. This approach is valid, however, as long as the domain wall network does not annihilate before EW symmetry breaking, nor at a temperature $T_{ann}$ slightly lower than $T_{ew}$. Since we assume that bias terms in the potential are very small, this assumption is valid, and the singlet domain wall network would only annihilate at a much later stage after EW symmetry breaking. Moreover, since the thermal corrections to the potential at temperatures of order $\mathcal{O}\text{(GeV)}$ would be negligible, using the zero-temperature scalar potential gives reliable results for the calculated profiles of the scalar fields. 

In the case when the bias terms (see equation (\ref{eq:treepot})) are not negligible, one needs first to verify the formation of domain walls at some temperature $T$ after the real singlet scalar field acquires a VEV. Then, we need to verify the existence of deeper vacua (since the presence of bias terms can significantly alter the potential) and also perform a precise study of the thermal evolution of the parameter points to determine whether the domain walls network annihilates before EW symmetry breaking making the domain walls harmless from the point of view of inducing the decay to deeper vacua. These aspects, however, are beyond the scope of the current study.

\begin{table}[h]
    \centering
    \begin{tabular}{ccccccccccccc}
         & $m_{h_1}$  & $m_{h_2}$  & $m_{h_3}$ & $m_{A}$  & $m_{H^{\pm}}$ & $\tan(\beta)$ & $v_s$  & $\alpha_1$ & $\alpha_2$ & $\alpha_3$ & $m_{12}$ \\
         \hline
         \hline
       $P_1$  & 95 & 125 & 616 & 743 & 609 & 2.54 & 1786 &  -0.37 & -1.49 & 0 & 359 \\
       $P_2$  & 125 & 410 &  733 & 481 & 368 & 2.11 & 4252 & 1.08 & -0.19 & -0.14 & 242 \\
       $P_3$  & 125 & 400 & 1200 & 400 & 470 & 1.88 & 1483 &  1.13 & -0.32 & -0.07 & 270 \\
       $P_4$  & 125 & 400 &  1200 & 279 & 429 &  2.93 & 2609 & 1.13& -0.16 & -0.12 & 198 \\
       \hline
       \hline
    \end{tabular}
    \caption{Benchmark parameter points with metastable long-lived EW vacua. The mass parameters $m_{h_1}$, $m_{h_2}$, $m_{h_3}$, $m_A$, $m_{H^{\pm}}$ as well as $v_s$ are given in GeV while $m_{12}$ is given in $\text{GeV}$.}
    \label{tab:benchamarks}
\end{table}

In the following, we discuss the vacuum decay via domain walls for some benchmark metastable parameter points (see Tab \ref{tab:benchamarks}) that have a very long lifetime. These benchmark points satisfy all theoretical and experimental constraints discussed in section \ref{appendixconstraints} and are phenomenologically viable since their decay rate to the deeper minimum is very small.  

\begin{figure}[h]
     \centering
      \subfloat[Evolution of $v_1(x)$]{\includegraphics[width=0.5\columnwidth]{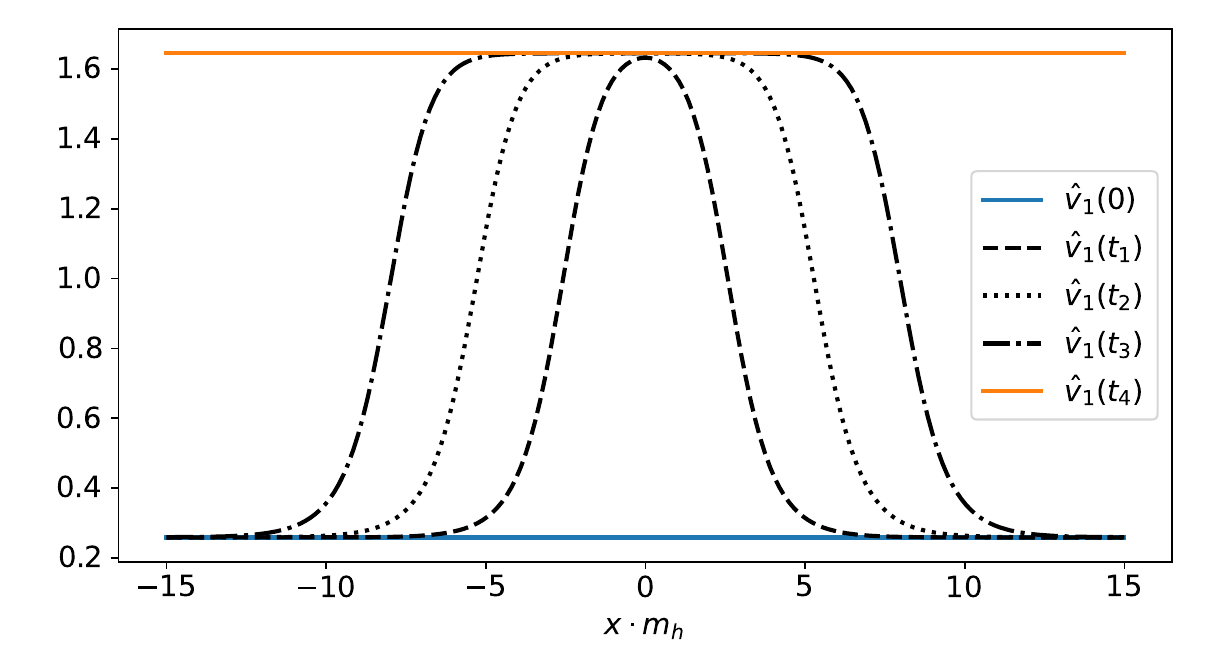}
      \label{subfig:v1decay}}
      \subfloat[Evolution of $v_2(x)$]{\includegraphics[width=0.5\columnwidth]{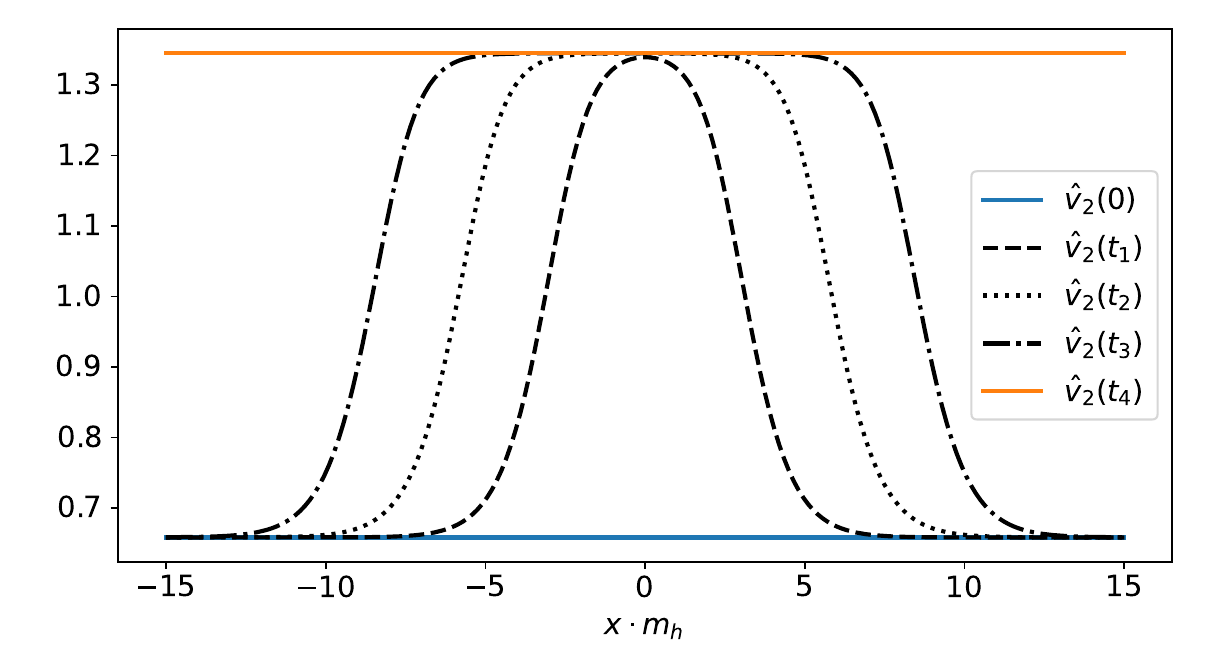}
       \label{subfig:v2decay}} \\
      \subfloat[Evolution of $v_s(x)$]{\includegraphics[width=0.5\columnwidth]{DomainWall/neutralvs.pdf}
      \label{subfig:vsdecay}}
\caption{Evolution of the normalized scalar fields $\hat{v}_1(x)$, $\hat{v}_2(x)$, and $\hat{v}_s(x)$ inside the domain wall (with $\hat{v}_i(x) = v_i(x)/v_{sm}$). We start at $t=0$ with the doublet scalar field configuration in the EW vacuum (shown in blue) and show the evolution of the scalar field configuration to the deeper vacuum (shown in orange). The scalar fields roll over to the deeper vacuum, and the wall of true vacuum expands outside, leading to the decay of the EW vacuum everywhere. The region with $\hat{v}_s=0$ expands in space, leading to the eventual decay of the domain wall. The time steps are given by: $t_1=22.5 m^{-1}_h$, $t_2=56.25 m^{-1}_h$, $t_3=90 m^{-1}_h$, and $t_4=191.25 m^{-1}_h$. } 
\label{fig:dwdecay}
\end{figure}

We show the results for the parameter point $P_1$ in Figure \ref{fig:dwdecay}. This parameter point leads to a global minimum for the potential of type $\mathcal{N}'$ that is deeper than the EW vacuum. For this parameter point \texttt{EVADE} gives a bounce action $B=95368$ for the tunneling rate (see equation (\ref{eq:tunnelingrate})) from the EW vacuum to the true vacuum of the potential. Therefore, the EW vacuum outside the wall is very long-lived and can be considered nearly stable, given its negligibly small decay rate. However, as is shown in Figure \ref{fig:dwdecay}, the Higgs doublets profiles inside the core of the domain wall rapidly change their values to correspond to the values of the deeper vacuum. This evolution occurs via a classical rollover from $(v_1,v_2,0)$ (shown in blue) to the true vacuum $(v'_1, v'_2,0)$ (shown in orange). Once this rollover is complete, the new vacuum nucleated inside the wall propagates outside of the wall since this expansion is energetically favorable, and the gain in energy via the expansion is much larger than the domain wall's tension that would otherwise stabilize the profile of the fields inside the wall\footnote{When considering thermal corrections, the transition and expansion of the deeper vacuum inside the region of the EW vacuum would only start when the gain in potential energy becomes larger than the domain wall's tension, this would then provide us with the temperature at which the decay occurs.}. 

In order to explain this behavior, we consider the effective potential of the Higgs doublets in the background of the singlet domain wall solution:
\begin{align}
 \notag  & V_{2HDM}(\Phi_1, \Phi_2, \Phi_s(x)) = \biggl(m^2_{11} + \frac{\lambda_7}{2}\Phi^2_s(x)\biggr) \Phi^{\dagger}_1\Phi_1 + \biggl(m^2_{22}+ \frac{\lambda_8}{2}\Phi^2_s(x)\biggr)\Phi^{\dagger}_2\Phi_2 \\ \notag &-  m^2_{12}(\Phi^{\dagger}_1\Phi_2 + h.c.)  + \frac{\lambda_1}{2}\bigl(\Phi^{\dagger}_1\Phi_1\bigr)^2 + \frac{\lambda_2}{2}\bigl(\Phi^{\dagger}_2\Phi_2\bigr)^2    + \lambda_3\bigl(\Phi^{\dagger}_1\Phi_1\bigr)\bigl(\Phi^{\dagger}_2\Phi_2\bigr) \\ 
    & + \lambda_4\bigl(\Phi_1^{\dagger} \Phi_2\bigr)\bigl(\Phi_2^{\dagger} \Phi_1\bigr) 
    +\biggl[\frac{\lambda_5}{2}\bigl(\Phi_1^{\dagger} \Phi_2\bigr)^2 + h.c\biggr]  + \frac{m^2_S}{2}\Phi^2_s(x) + \frac{\lambda_6}{8}\Phi^4_s(x) .
\end{align}

\begin{figure}[H]
     \centering
     \subfloat[2HDM potential outside the wall]{\includegraphics[width=0.5\columnwidth]{DomainWall/P1potentialoutside.pdf}
         \label{subfig:outsidewall}}
     \subfloat[2HDM potential inside the wall]{\includegraphics[width=0.5\columnwidth]{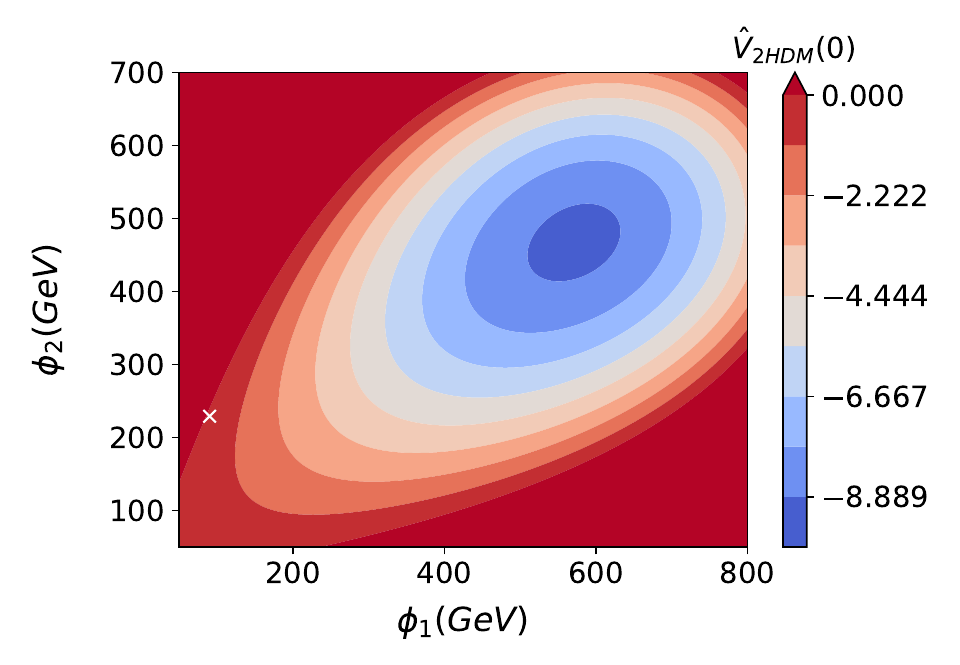}
      \label{subfig:insidewall}}
\caption{2HDM potential in the background of the singlet domain wall $v_s(x)$. (a) 2HDM potential outside the wall, the EW vacuum is represented with a white cross and is protected from decaying to deeper vacua due to the presence of the potential barrier. (b) 2HDM potential inside the wall ($v_s=0$). The EW vacuum is represented by a white cross. Inside the wall, the barrier between the EW vacuum and the deeper vacuum vanishes, and the scalar fields roll over from the EW vacuum to the deeper vacuum.} 
\label{fig:2hdmpot}
\end{figure}

We show in Figure \ref{fig:2hdmpot} the potential of the Higgs doublets for $P_1$ in the background of the domain wall, both outside (left) and inside the wall (right). The EW vacuum is represented by a white cross, and while it is protected by the potential barrier from tunneling to deeper vacua, the barrier disappears inside the wall, and the EW vacuum can roll over to the deeper vacuum. Notice that, inside the wall, $(v_1,v_2,0)$ is not a stationary point of the potential and therefore the field will be unstable at that point.   

\begin{figure}[h]
     \centering
         \subfloat[Evolution of $v_1(x)$]{\includegraphics[width=0.5\columnwidth]{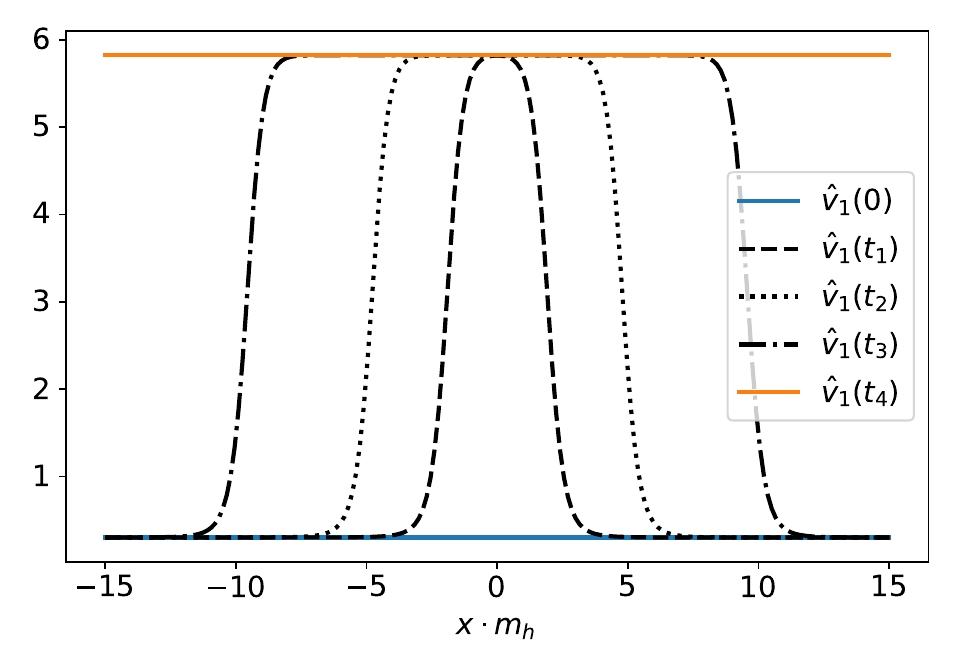}
         \label{subfig:v1CBdecay}}
         \subfloat[Evolution of $v_+(x)$]{\includegraphics[width=0.5\columnwidth]{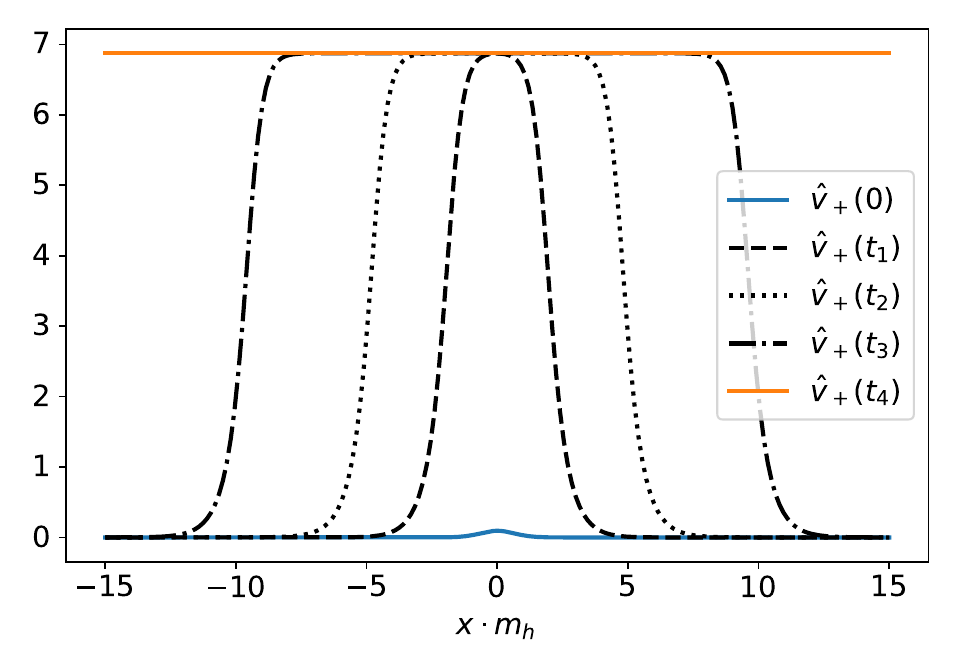}
         \label{subfig:vplusCBdecay}} \\
         \subfloat[Evolution of $v_s(x)$]{\includegraphics[width=0.5\columnwidth]{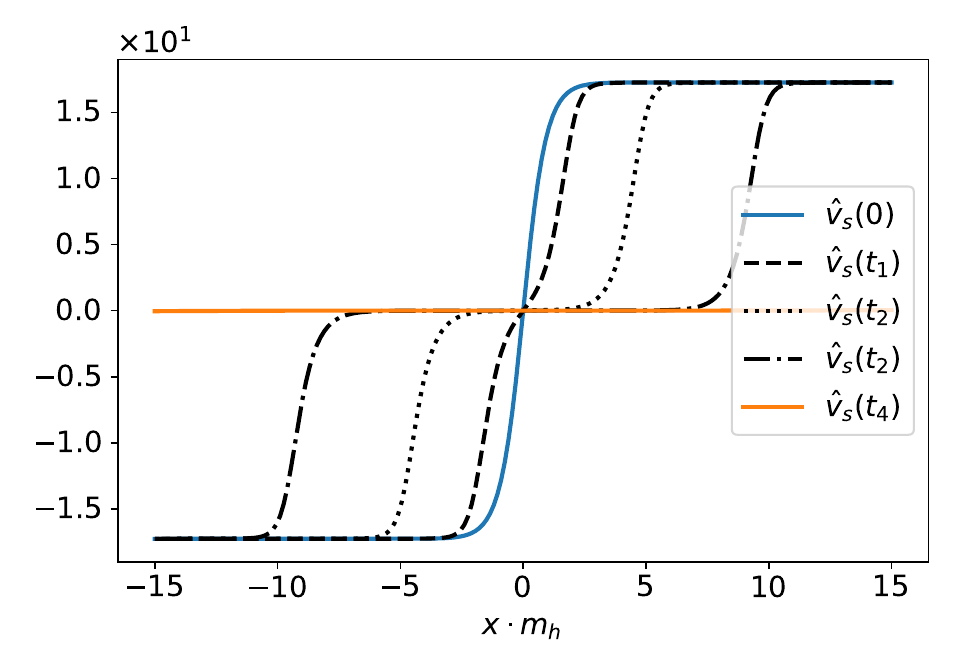}
         \label{subfig:vsCBdecay}}
         \subfloat[Evolution of $v_2(x)$]{\includegraphics[width=0.5\columnwidth]{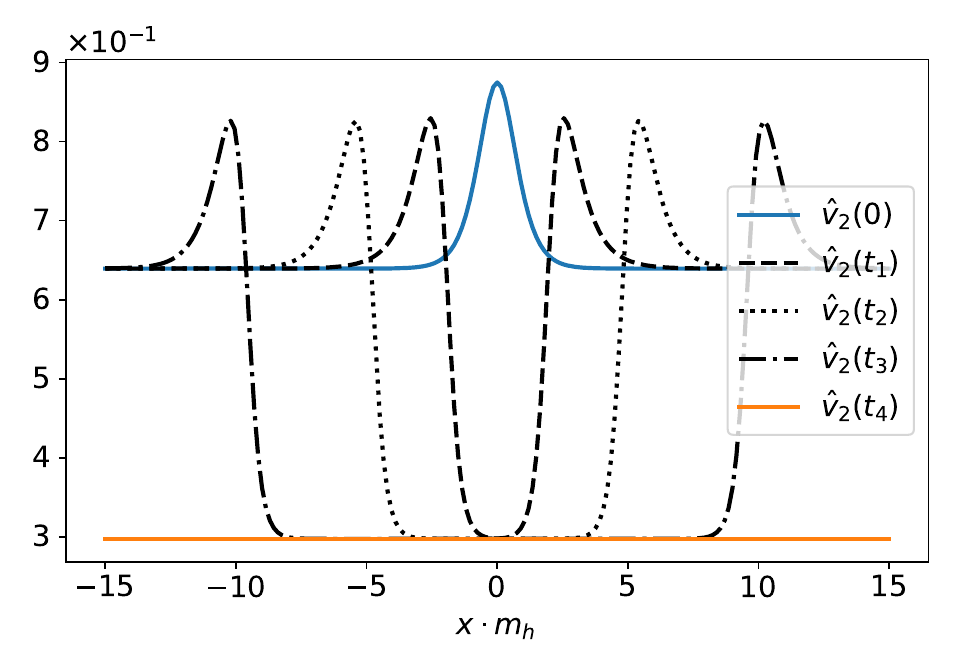}
        \label{subfig:v2CBdecay}}
\caption{Evolution of the normalized scalar fields configuration $\hat{v}_i(x) = v_i(x)/v_{sm}$ inside the domain wall for the case when the deeper minimum is of $\mathcal{CB}$ type (benchmark point $P_2$). We start with the doublet scalar fields in the EW vacuum with a small fluctuation in $v_+(0)$ (shown in blue) and plot the evolution to the deeper vacuum (shown in orange). In this case, $v_+(0)$ acquires a VEV inside the wall, which then expands everywhere. The time steps are given by: $t_1=10.25 m^{-1}_h$, $t_2=28.12 m^{-1}_h$, $t_3=56.25 m^{-1}_h$, and $t_4=90 m^{-1}_h$.} 
\label{fig:dwdecaycharged}
\end{figure}

The deeper vacuum can also be of electric charge breaking type $\mathcal{CB}$ or CP-breaking type $\mathcal{CP}$ (e.g. $P_2$ and $P_4$ in Table \ref{tab:benchamarks}). We verified that the rollover transition of the EW to such deeper vacua also occurs. For the case of $\mathcal{CB}$ global minimum (shown in Figure \ref{fig:dwdecaycharged}), small values of $v_+(0)$ start growing inside the wall, reach the values of a deeper vacuum, and then expand outside the wall. Here again, the barrier between the EW vacuum and the global minimum disappears inside the domain wall, and the decay occurs via a classical rollover to the $\mathcal{CB}$ vacuum as shown in Figure \ref{fig:dwdecaycharged}. 

\begin{figure}[h]
     \centering
        \subfloat[Evolution of $v_1(x)$]{\includegraphics[width=0.49\textwidth]{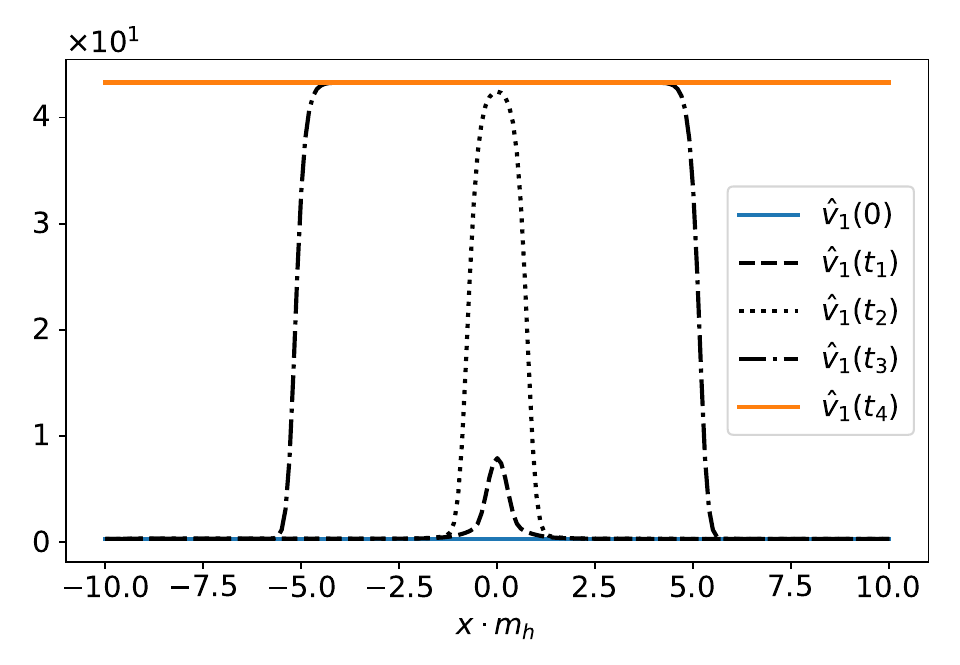}
         \label{subfig:v1CPdecay}}
        \subfloat[Evolution of $v_2(x)$]{\includegraphics[width=0.49\textwidth]{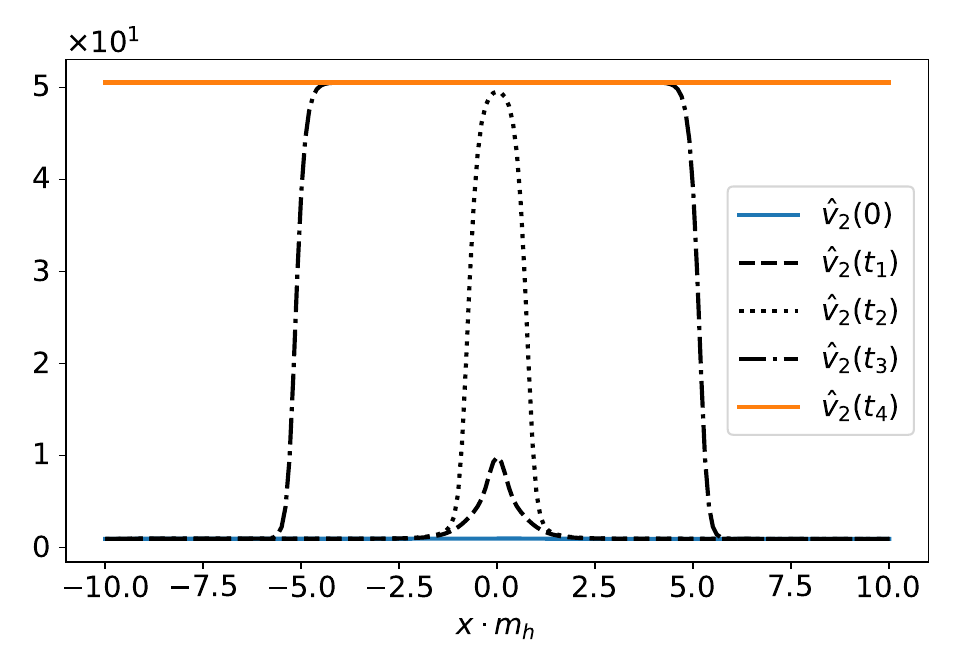} 
        \label{subfig:v2CPdecay}}  \\
        \subfloat[Evolution of $v_s(x)$]{\includegraphics[width=0.49\textwidth]{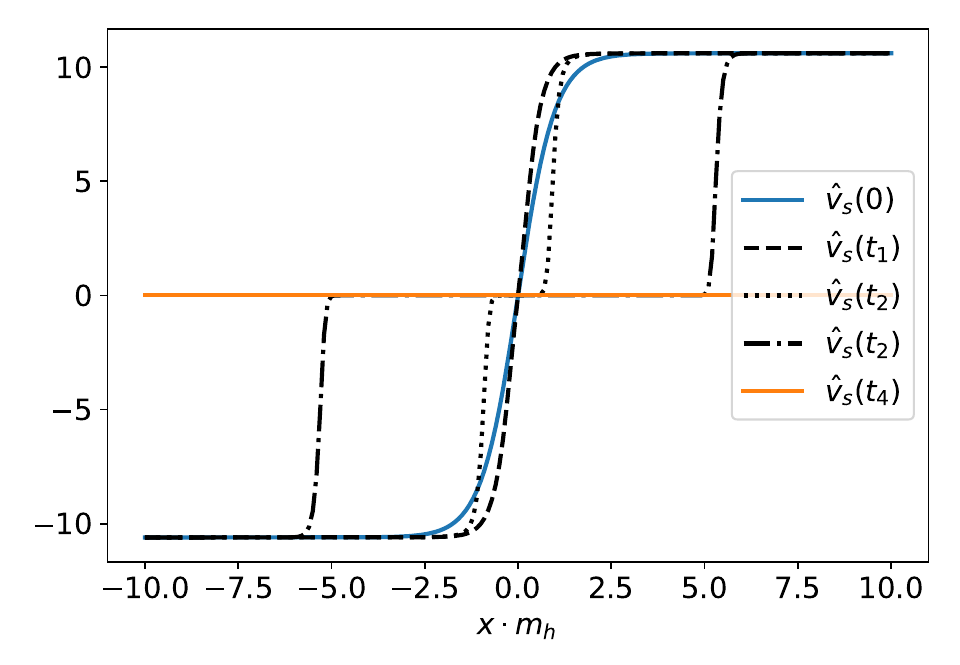}
         \label{subfig:vsCPdecay}}
        \subfloat[Evolution of $\xi(x)$]{\includegraphics[width=0.49\textwidth]{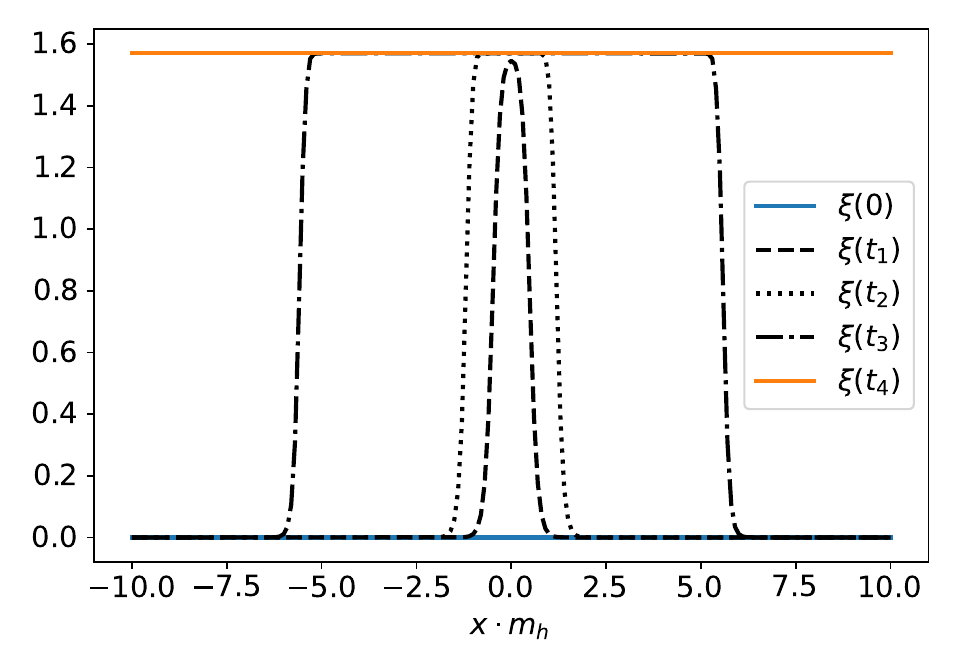}
         \label{subfig:xidecay}}
\caption{Evolution of the scalar fields inside the domain wall For the case when the deeper minimum is of $\mathcal{CP}$ type. We start with the doublet scalar fields in the EW vacuum and show the evolution to the deeper vacuum. In this case, $\xi(x=0)$ acquires a non-vanishing value inside the wall, which then expands everywhere. The time steps are given by: $t_1=10 m^{-1}_h$, $t_2=20 m^{-1}_h$, $t_3=50 m^{-1}_h$, and $t_4=100 m^{-1}_h$.} 
\label{fig:dwdecaycp}
\end{figure}

A similar behavior (see Figure \ref{fig:dwdecaycp}) occurs for the decay of the EW metastable vacuum into a CP-violating global vacuum. The CP-violating phase $\xi(0)$ grows inside the wall, and once the global minimum is nucleated, it quickly expands outside. 

These benchmark points illustrate the possibility of using the mechanism of vacuum decay induced by domain walls in order to rule out parameter points with long-lived EW minima. In all of these cases, the 2HDM potential inside the core of the domain wall, where $v_s=0$, only has a single minimum. In the next subsection, we look into the case when this 2HDM potential develops several minima.    

\subsection{Several extrema inside the wall}
\begin{figure}
    \centering
    \includegraphics[width=\linewidth]{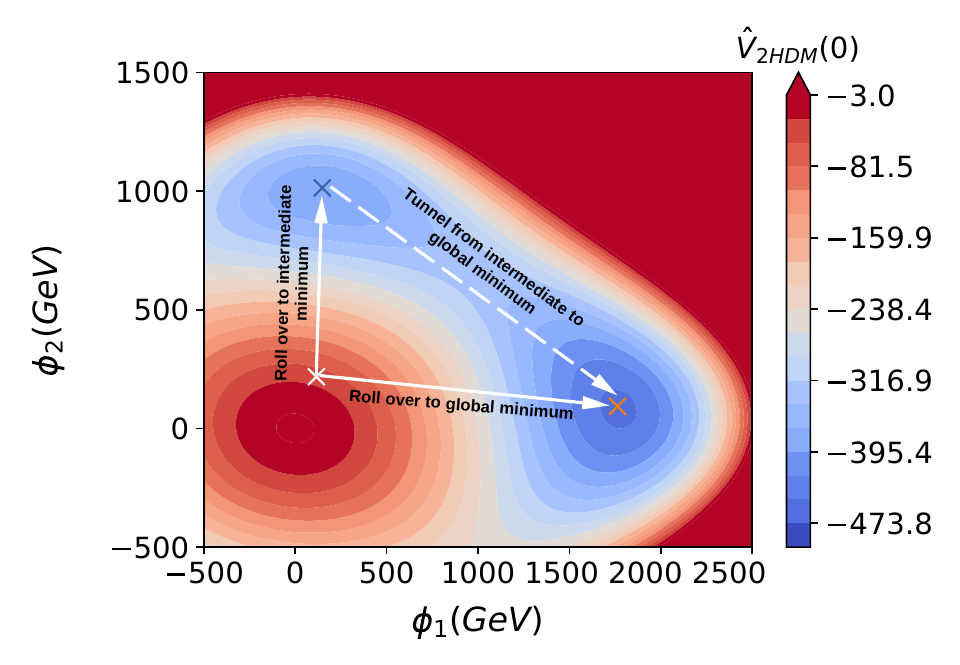}
    \caption{2HDM potential inside the wall for parameter point $P_3$. The initial field values $(v_1, v_2)$ are shown in a white cross mark. We find two minima for the 2HDM potential, with the intermediate minimum (blue cross mark) being separated from the global minimum (orange cross mark) by a potential barrier. }
    \label{fig:severalminima}
\end{figure}
For the discussed benchmark points, we observed that the decay of the EW vacuum inside the DW happens via a classical rollover into the true global vacuum of the potential. This happens due to the disappearance of the potential barrier between the initial field configuration $(v_1,v_2,0)$ and the true global minimum $(v^{true}_1, v^{true}_2,0)$. Inside the wall, the 2HDM part of the EW vacuum $(v_1,v_2,0)$ is not an extremum of the 2HDM potential at $x=0$ (recall that inside the wall $v_s=0$) since it does not, in general, satisfy the minimization conditions of the 2HDM potential in the $\phi_1$ and $\phi_2$ directions:
\begin{align}
    & m^2_{11} + \dfrac{1}{2}(\phi^2_1\lambda_1 + \phi^2_2 \lambda_{345}) - m^2_{12}\dfrac{\phi_2}{\phi_1} = 0, \\
    & m^2_{22} + \dfrac{1}{2}(\phi^2_1\lambda_1 + \phi^2_2 \lambda_{345}) - m^2_{12}\dfrac{\phi_1}{\phi_2} = 0.
\end{align}
We therefore have two possible scenarios:

\noindent
1) The 2HDM potential at $x=0$ has only a single minimum that corresponds to the true global minimum. In this case, since $(v_1,v_2,0)$ is not a stationary point of the potential, the EW vacuum experiences a rollover to the true minimum, and we obtain a classical vacuum decay induced by the domain wall.

\noindent
2) The 2HDM potential at $x=0$ has at least another intermediate minimum $(v^{int}_1,v^{int}_1,0)$ between $(v_1,v_2,0)$ and the true global minimum of the potential. In this case, several scenarios may occur (see Figure \ref{fig:severalminima}):
\begin{itemize}
    \item The fields roll over direclty to the global minimum (shown in an orange cross mark in Figure \ref{fig:severalminima}).
    \item The fields roll over from the EW minimum configuration $(v_1,v_2)$ to $(v^{int}_1, v^{int}_2)$ (shown in a blue cross mark) and are trapped there, due to the potential barrier between the intermediate minimum and the global minimum. 
    \item The fields roll over to the intermediate minimum and later tunnel to the true minimum.
\end{itemize}
In the case when the intermediate minimum is lower in the potential than the EW minimum $(v_1,v_2,v_s)$, we will also experience a classical rollover vacuum decay via domain walls into this intermediate vacuum, even if we don't reach the true global minimum of the potential. The possibility of tunneling inside the DW from the intermediate vacuum into the true global vacuum requires a detailed study of the temperature evolution of the effective potential as well as the annihilation time of the domain wall network, and is beyond the scope of our work. We therefore reserve its investigation for a future study\footnote{In all the parameter scans that will be discussed later, we didn't find a parameter point where the fields are trapped in an intermediate minimum, under the assumption that the friction term d in (\ref{eq:eom}) is small.}. This scenario was studied in the real scalar singlet extension of the SM \cite{Blasi:2022woz, Agrawal:2023cgp} in the context of vacuum trapping, where the DW facilitates the tunneling from the symmetric vacuum with $v_h = 0$ to the electroweak vacuum $v_{sm} \approx 246 \text{ GeV}$.     

The existence of several distinct minima for the 2HDM potential inside the wall can complicate the analysis of the fate of the EW metastable minimum. When intermediate minima exist (see e.g. the potential in Figure \ref{fig:severalminima}), one needs to verify whether the EW vacuum decays or not by following the time evolution of the scalar fields inside and outside the wall. This also depends heavily on the magnitude of the friction term $d$ in (\ref{eq:eom}), since the field configuration would oscillate around the intermediate minimum and can eventually move over the potential barrier to the global minimum. If, however, the true global vacuum is the only minimum of the 2HDM potential inside the DW, then the EW vacuum will necessarily decay to the true vacuum via a classical rollover, and no DW simulation is needed. 

It is possible in the 2HDM to determine analytically if the potential has several coexisting stationary points \cite{Barroso:2013awa,Barroso:2012mj}. In those publications, the necessary conditions\footnote{These are necessary and sufficient conditions for the existence of four stationary points, but are not sufficient conditions to ensure that two of these stationary points are indeed different minima.} needed so that the 2HDM potential can have two coexisting neutral minima have been discussed. These two minima are the EW vacuum and another one dubbed the panic vacuum, which could lie deeper in the potential and lead to different masses of the gauge bosons and SM fermions. The two conditions for the possibility of the presence of two different neutral minima in the 2HDM potential are:
\begin{align}
   & m^2_{22} + k^2m^2_{22} < 0, \\
   & \sqrt[3]{x^2} + \sqrt[3]{y^2} \le 1,
\end{align}
where:
\begin{align}
   & x = \dfrac{4km^2_{12}}{m^2_{11}+k^2m^2_{22}} \:  \dfrac{\sqrt{\lambda_1\lambda_2}}{\lambda_{345}-\sqrt{\lambda_1\lambda_2}},\\
   & y = \dfrac{m^2_{11}-k^2m^2_{22}}{m^2_{11}+k^2m^2_{22}} \: 
 \dfrac{\lambda_{345}+\sqrt{\lambda_1\lambda_2}}{\sqrt{\lambda_1\lambda_2}-\lambda_{345}}, \\
   & k = \sqrt[4]{\dfrac{\lambda_1}{\lambda_2}}.
\end{align}
If these conditions are not met, the 2HDM potential only admits a single minimum, and we conclude that the DW will necessarily induce the rollover transition of the EW vacuum. However, if these conditions are met, we verify the existence of intermediate minima in the 2HDM potential at $x=0$ using \texttt{EVADE}. For parameter points where the extra stationary points are either maxima or saddle points, the EW vacuum decays via a classical rollover to the deeper vacuum. However, suppose the extra stationary points are local minima. In that case, one has to numerically verify the real-time evolution of the scalar field configurations inside the DW to determine the fate of the EW metastable vacuum. 
\begin{figure}[t]
\centering
\includegraphics[width=0.75\linewidth]{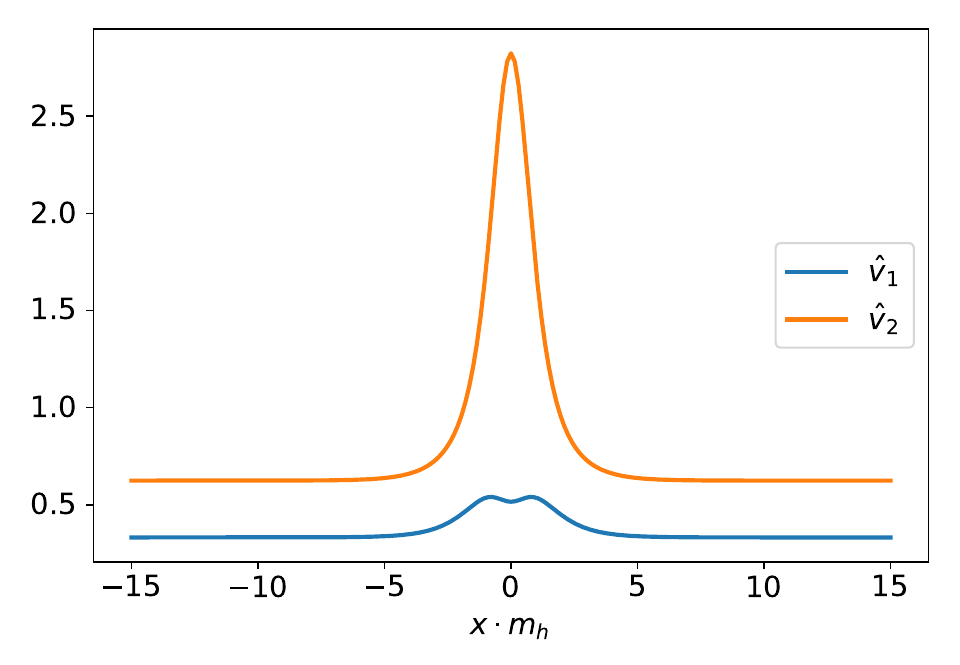}
\caption{DW solutions of parameter point $P_3$ with $\hat{d}=\frac{d}{m_h} = 1$, and $\hat{v}_i = v_i/v_{sm}$. In this case, the friction term is large enough to dissipate the kinetic energy in the scalar field, causing them to be trapped in the intermediate minimum. } 
\label{fig:dwp3}
\end{figure}
We consider $P_3$ (see Table \ref{tab:benchamarks}) as an example for a parameter point where the 2HDM potential inside the wall has at least one other minimum alongside the deepest global minimum. The local minimum at $(v_{1,i},\text{ } v_{2,i}, v_{s,i}) = (172\text{ GeV},\text{ } 991\text{ GeV}, 0)$ lies higher in the potential than the EW vacuum $(v_1,v_2,v_s)$. Notice that there is no barrier between the starting field values $(v_1,v_2,0)$ (shown as a white cross mark in Figure \ref{fig:severalminima}) and the two minima, while there is a potential barrier between both minima.

In this case, the fate of the electroweak vacuum depends on the value of the friction term $d$ in (\ref{eq:eom}). For $\hat{d}=\frac{d}{m_h} = 1$ (see Figure \ref{fig:dwp3}), where $\hat{d}$ is the dimensionless friction term\footnote{We make all quantities in the equation of motion (\ref{eq:eom}) dimensionless when solving it numerically.}, the EW vacuum rolls over to the intermediate minimum of the 2HDM potential inside the wall, briefly oscillates around that minimum, and gets trapped in that vacuum. This occurs since the kinetic energy of the scalar fields is not large enough to cross the barrier to the global minimum, due to the friction term dissipating it. This leads the EW vacuum to remain metastable (since the EW vacuum outside the wall lies lower in the N2HDM potential than this intermediate minimum). However, the potential barrier between the intermediate minimum inside the wall and the global minimum can be smaller than the potential barrier between the metastable EW minimum and the global minimum, one should therefore consider the possibility of quantum tunneling inside the wall. These scenarios were studied in detail in the case of the real singlet scalar extension of the SM, see e.g. \cite{Agrawal:2023cgp,Blasi:2022woz}, where the domain walls have been used as impurities to seed the EW phase transition inside the wall and to rescue parameter points from vacuum trapping. 

If the friction term gets smaller e.g. $\hat{d}=0.05$, we observe the classical rollover to the intermediate minimum, the field oscillates around that minimum, and then crosses the barrier to the global minimum. Once the global minimum is nucleated inside the wall, we obtain the same behavior observed for the usual EW vacuum decay via domain walls. 

Since we are working in the zero temperature limit at the radiation domination era, the Hubble parameter $\mathcal{H} \propto T^2/M_{pl}$ \cite{Kolb:1990vq}, which would act as a component of the friction term $d$, is very suppressed. Another, a priori, more sizable contribution might come from the interaction of the scalar fields with the thermal plasma, leading to the damping of the kinetic energy that the doublet fields start with. The evaluation of this contribution depends heavily on the thermal evolution and the coupling between the scalars and fermions.

It is also crucial to determine whether this intermediate minimum is present immediately after the EW phase transition, or whether it develops over time. In case the global minimum is the only minimum present in the 2HDM potential inside the wall at high temperatures after EW phase transition, the classical rollover will occur to the global minimum, leading to the decay of the EW vacuum. However, in case the intermediate minimum is also present in the potential at high temperatures just after EW phase transition, it is then possible that, after the EW phase transition, the intermediate vacuum is trapped inside the wall. This could happen if the doublet fields lose their kinetic energy during their roll over. As the universe cools down, the friction term gets smaller, and as the singlet domain wall moves to other regions of the universe, the doublet fields experience less friction, and those regions might undergo a domain wall-induced EW vacuum decay.

In our examples, we saw that, inside the wall, there is no barrier between the initial doublet field configuration $(v_1, v_2, 0)$ and the two minima of the 2HDM potential (the global minimum and the intermediate minimum). Thermal or quantum fluctuations can also play an important role in determining in which direction the doublet fields start rolling over. All these aspects require a careful study of the thermal history of particular parameter points, and we leave their further discussion for a future study. 

\section{Phenomenological Scenarios for Metastability}
\label{section5}
\begin{table}[h]
    \centering
    \begin{tabular}{ccccccccc}
         & $m_{H^{\pm}}$  & $m_{A}$ & $v_s$  & $C^2_{h_2VV}$ & $C^2_{h_2t\bar{t}}$ & $R_{23}$ & $m^2_{12}$ \\
         \hline
         \hline
       S1  & [200, 1500]  & [200, 1500] & [1, $10^4$] & [0.6, 1] &  [0.6, 1] & [-1, 1] & [0, 2$\cdot 10^5$]  \\
        \hline
       & $m_{H^{\pm}}$  & $m_{A}$  & $v_s$  & $C^2_{h_1VV}$ & $C^2_{h_1t\bar{t}}$ & $R_{13}$ & $m^2_{12}$ \\
         \hline
         \hline
       S2  & [200, 1500]  & [200, 1500] & [1, $10^4$] & [0.6, 1] &  [0.6, 1] & [-1, 1] & [0, 5$\cdot 10^4$]  \\
       \hline
       \hline
    \end{tabular}
    \caption{Set of input parameters for our \texttt{ScannerS} scans with all parameters in GeV unit, except $m^2_{12}$ which is given in $\text{GeV}^2$. For Scenario 1 (S1), the masses of the CP-even Higgs particles are: $m_{h_1} \in [94,96] \text{ GeV}$, $m_{h_2} = 125.09 \text{ GeV}$ and $m_{h_3} \in [200, 1500]\text{ GeV}$. For Scenario 2 (S2), the masses of the CP-even Higgs particles are: $m_{h_1} = 125.09\text{ GeV}$, $m_{h_2} = 400\text{ GeV}$ and $m_{h_3} \in \{600, 800, 1200\}\text{ GeV}$. For both scans, the range of $v_s$ is between 1 and $10^4$ GeV, while $\tan(\beta) \in [0.5,3]$. The scans were conducted over all types 1-4 (see Table \ref{Tab:2hdmtypes}).}
    \label{tab:scans}
\end{table}
In this section, we discuss some phenomenological scenarios leading to the separation of parameter regions into metastable and stable EW vacua. This is done in order to pinpoint regions of parameter space where EW vacuum decay via domain walls can occur. We saw in the previous section that a free parameter scan did not provide an obvious correlation between parameter variables and the nature of the EW vacuum. We now study some specific scenarios where we fix some parameter variables of the model. 

For the parameter scans (generated using \texttt{ScannerS}), we impose the theoretical constraints of boundedness from below and perturbative unitarity (see Appendix \ref{appendixconstraints}). We also make sure that the generated parameter points are not unstable i.e. all generated parameter points are either stable or have a bounce action $B>390$. We don't directly impose the requirement of $Z'_2$ and electroweak symmetry restoration in the early universe (based on the analytical conditions derived in \cite{Biekotter:2021ysx} and summarized in Appendix \ref{appendixconstraints}). However, we later check if those conditions are fulfilled by the generated set of parameter points. These analytical conditions for symmetry restoration are based on the Arnold-Espinosa daisy resummation method. It was argued in \cite{Bittar:2025lcr} that other resummation methods do not show symmetry non-restoration in the 2HDM. Since this might also be the case in the N2HDM, we also include, in our analysis, parameter points that feature the possibility of symmetry non-restoration. This is done in order not to overlook other possible metastable parameter points that would be ruled out by vacuum decay via domain walls and, more importantly, not to overlook parameter regions that have a stable EW vacuum but feature symmetry non-restoration. If, however, symmetry non-restoration actually occurs (up to ultra-high temperatures in the early universe to avoid the formation of domain walls) and is not an artifact of the resummation method, then these parameter points could still be rescued from vacuum decay via domain walls. In any case, a careful determination of the thermal evolution of the scalar potential at high temperatures using different resummation schemes might be crucial in order to determine whether the $Z'_2$ symmetry is restored or not in the early universe. 

In case when parameter points with symmetry-restored metastable EW vacua overlap with symmetry non-restored metastable parameter points in the whole scan range, we plot both types in orange. Metastable parameter points with symmetry non-restoration are only shown explicitly if they can be separated from regions with only symmetry restoration. Therefore, unless otherwise specified, parameter regions in orange include both types of metastable vacua.

We start our discussion with scans where the experimental constraints of electroweak precision observables, as well as flavor and Higgs searches, are not yet imposed. Since we want to emphasize the possibility of our mechanism in inducing the EW vacuum decay via domain walls as a new important constraint on its own, it is important to make sure that experimental constraints don't filter parameter points with stable EW vacua from regions that also have metastable vacua. We discuss in section \ref{experimental} a scan where all experimental constraints were imposed in addition and show that the mechanism of EW vacuum decay via domain walls could rule out parameter regions that are otherwise still viable.

\subsection{Scenario 1: a 95 GeV CP-even Higgs particle}
\subsubsection{Scan without applying experimental constraints}

\begin{figure}[h]
     \centering
     \subfloat[]{\includegraphics[width=0.5\columnwidth]{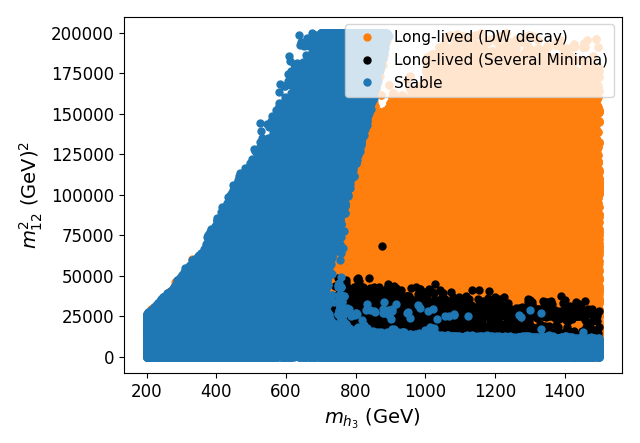}
       \label{subfig:scan1mh3m12}}
     \subfloat[]{\includegraphics[width=0.5\columnwidth]{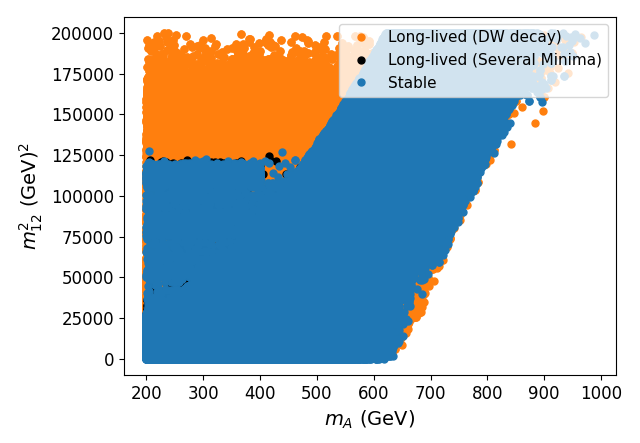}
      \label{subfig:scan1mhcm12}} \\
    \subfloat[]{\includegraphics[width=0.5\columnwidth]{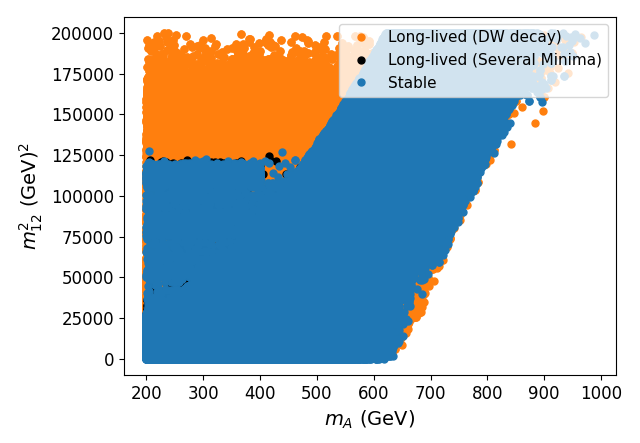}
    \label{subfig:scan1mham12}}
\caption{Results of Scan 1 showing parameter points with a metastable EW minimum with a single minimum inside the domain wall (in orange), a metastable EW minimum with several minima inside the wall (in black), and a stable EW minimum (in blue) plotted on top of the metastable points. The results of this scan demonstrate the possibility of separating the metastable-only region from the regions of parameter space that include both stable and metastable vacua.}
\label{fig:scenario1noexp}
\end{figure}

We first start with a general parameter scan where we take the lightest CP-even Higgs boson to have a mass between $94 \text{ GeV}$ and $96 \text{ GeV}$, which is motivated by the excesses observed by the CMS \cite{CMS:2024yhz} and ATLAS collaborations \cite{ATLAS:2024itc}. We fix the mass of the CP-even Higgs particle $h_2$ to $m_{h_2}=125.09 \text{ GeV}$, corresponding to the SM-like Higgs boson, and allow the mass of $h_3$ to vary between $200 \text{ GeV}$ and $1500 \text{ GeV}$. We vary the masses of the charged and CP-odd Higgs bosons as well as the parameters $m^2_{12}$, $\tan{\beta}$, $v_s$, and the mixing angles (see Table \ref{tab:scans}).

The results are shown in Figure \ref{fig:scenario1noexp}. The parameter points with a stable electroweak minimum are shown in blue, while the parameter points with a metastable EW minimum and a global minimum with $v_s = 0$ are shown in orange. We also plot (in black) parameter points with a metastable EW minimum where at least one extra minimum other than the global minimum is found for $v_s=0$. The metastable parameter points are plotted first and then the stable ones on top\footnote{This is done to separate stable and metastable regions of parameter space. Blue regions also feature a large number of metastable parameter points that would be excluded individually.}. We observe a clear separation between stable and metastable regions only in the $m_{h_3}-m^2_{12}$ as well as the $m_{A}-m^2_{12}$, and the $m_{H^{\pm}}-m^2_{12}$ planes. This metastable-only region is obtained for $m_{h_3} > 900 \text{ GeV}$ and $m^2_{12} > 30000 \text{ GeV}$. The obtained results do not show any dependence on $v_s$. The dependence on the mixing angles is modeled in our case by the singlet admixture $\Sigma_i = R^2_{i3}$ as well as the coupling of the SM-like Higgs boson to gauge bosons ($c^2_{h_2VV}$) and top quarks ($c^2_{h_2t\bar{t}}$). Both stable and metastable minima overlap over the scanned parameter regions for the mixing angles. Note that the symmetry non-restoration metastable points overlap with the ones where we obtain symmetry restoration. Therefore, one could exclude those regions based on DW induced vacuum decay only if symmetry non-restoration is an artifact of the Arnold-Espinosa daisy resummation scheme.

Motivated by these results, we perform several case studies of specific parameter scans:
\begin{enumerate}
    \item Scan of the region $1000 \text{ GeV} < m_{h_3} < 1500 \text{ GeV}$ and $(2.5 \cdot 10^4 < m^2_{12}<5 \cdot 10^5) \text{ GeV}^2$. 
    \item Scan with fixed $m^2_{12} = 2\cdot 10^4 \text{ GeV}^2$ and variable $m_{h_3}$, $\tan(\beta)$, $m_{A}$, and $m_{H^{\pm}}$.
    \item Scan with fixed $m_{h_3} = 600 \text{ GeV}$ and variable $ m_{H^{\pm}}$, $m_{A}$, $\tan(\beta)$, and $m^2_{12}$.
    \item Scan with fixed $m_{h_3} = 1000 \text{ GeV}$ and variable $ m_{H^{\pm}}$, $m_{A}$, $\tan(\beta)$, and $m^2_{12}$.
    \item Scan with fixed $m_{h_3} = 1000 \text{ GeV}$ and variable $ m_{H^{\pm}}$, $m_{A}$, and $m^2_{12}$ but in the alignment limit ($c^2_{h_2VV}= c^2_{h_2t\bar{t}}=1$).
\end{enumerate}

\begin{figure}[h]
	\centering
	\subfloat[]{\includegraphics[width=0.5\columnwidth]{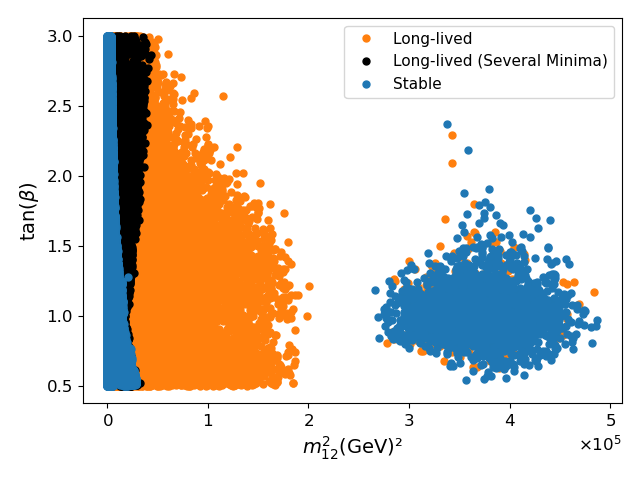}
		\label{subfig:scan1tanbetam12}}
	\subfloat[]{\includegraphics[width=0.5\columnwidth]{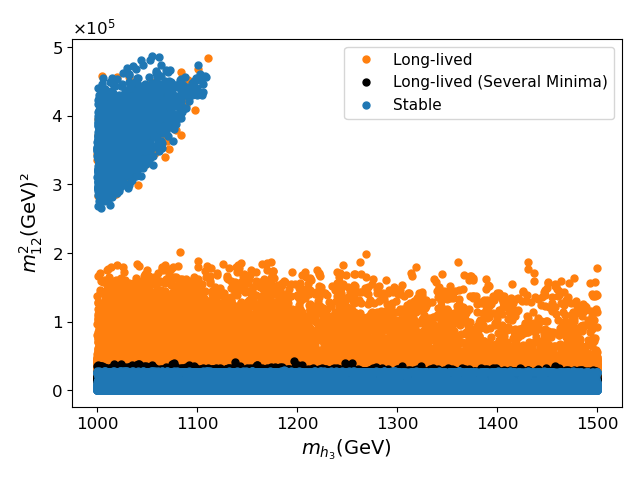}
		\label{subfig:scan1tanbetam3}} \\
	\subfloat[]{\includegraphics[width=0.5\columnwidth]{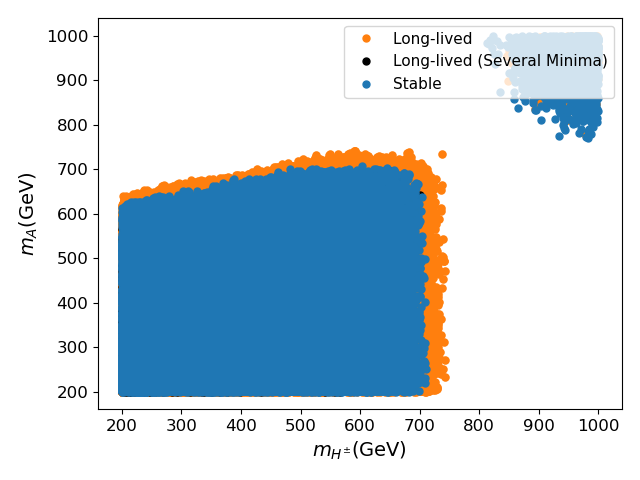}
		\label{subfig:scan1mcma}}
	\subfloat[]{\includegraphics[width=0.5\columnwidth]{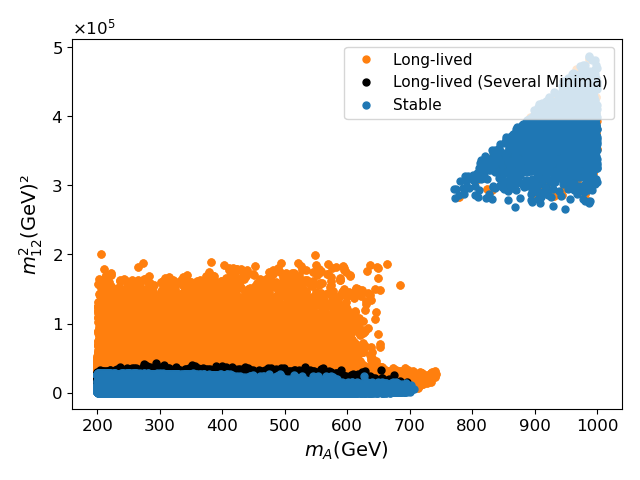}
		\label{subfig:scan1mam12}}
	\caption{Results of Scan 1 showing the possibility to separate the metastable-only region from the regions of parameter space which include both stable and metastable vacua. We also show (in black) the metastable parameter points with several minima at $v_s=0$.}
	\label{fig:scenarios1}
\end{figure}

The results for scan 1 are shown in Figure \ref{fig:scenarios1}. We find that the region of metastability depends heavily on $m^2_{12}$, and, to a lesser extent, on $\tan(\beta)$. The region of vacuum stability is concentrated at $m^2_{12} < 3 \cdot 10^4 \text{ GeV}^2$ for $\tan(\beta) < 1$. For higher values of $\tan(\beta)$ the region of vacuum stability is shifted to $m^2_{12}<10^4 \ \text{GeV}^2$. In scan 1, we find that all parameter points between $ (3\cdot 10^4 < m^2_{12} < 2\cdot 10^5) \ \text{GeV}^2$ are metastable. Note that for $m^2_{12} >> 2.5 \cdot 10^5 \text{ GeV}^2$, we also find a region where vacuum stability and metastability coexist.  

\begin{figure}[H]
	\centering
	\subfloat[]{\includegraphics[width=0.5\columnwidth]{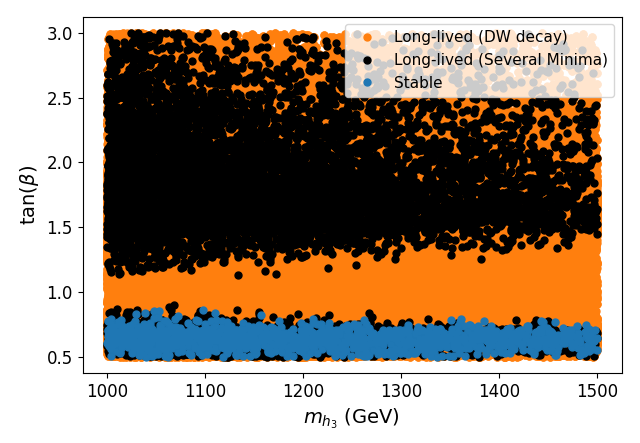}
		\label{subfig:scan2tanbetam3}}
	\subfloat[]{\includegraphics[width=0.5\columnwidth]{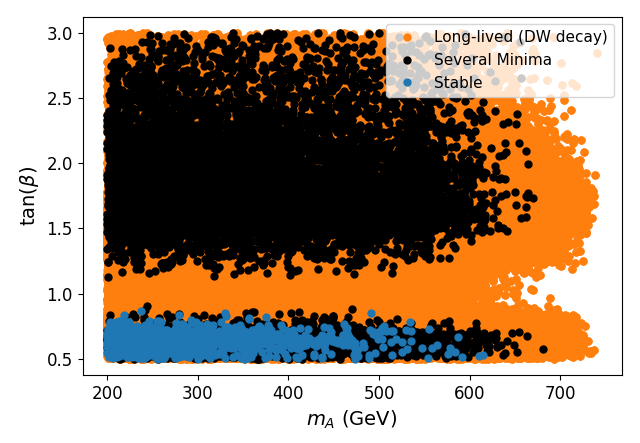}
		\label{subfig:scan2mhatanb}} 
	\caption{Results of Scan 2 showing the possibility to separate the metastable-only region from the regions of parameter space which include both stable and metastable vacua. We find that parameter points with stable EW vacua are concentrated at low $\tan(\beta)$. }
	\label{fig:scan2}
\end{figure}
\begin{figure}[H]
	\centering
	\subfloat[]{\includegraphics[width=0.5\columnwidth]{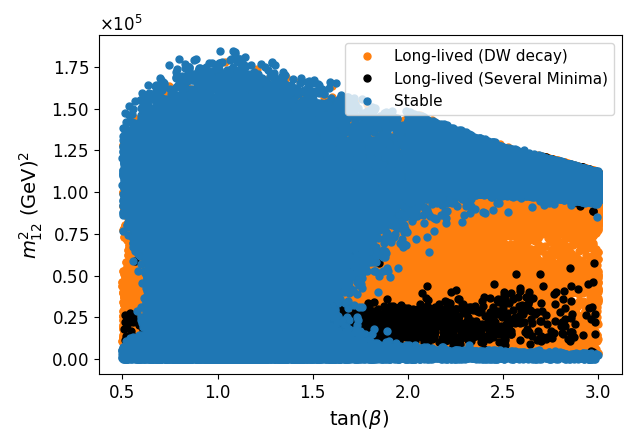}
		\label{subfig:scan3tanbetam12}}
	\subfloat[]{\includegraphics[width=0.5\columnwidth]{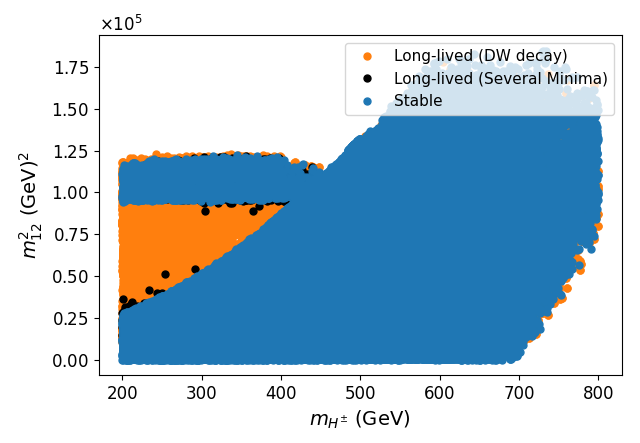}
		\label{subfig:scan3mhcm12}} 
	\caption{Results of Scan 3 showing the possibility to separate the metastable-only region from the regions of parameter space which include both stable and metastable vacua.}
	\label{fig:scan3}
\end{figure}
\begin{figure}[H]
	\centering
	\subfloat[]{\includegraphics[width=0.5\columnwidth]{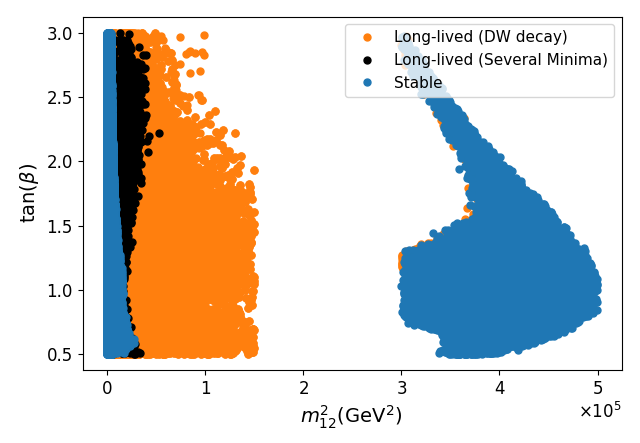}
		\label{subfig:scan4tanbetam12}}
	\subfloat[]{\includegraphics[width=0.5\columnwidth]{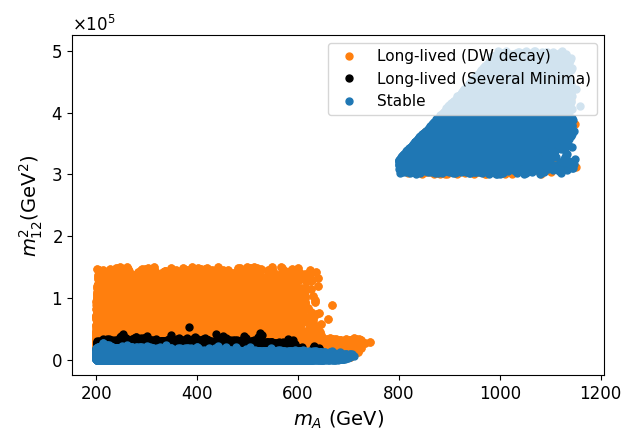}
		\label{subfig:scan4mhcmha}} 
	\caption{Results of Scan 4 showing the possibility to separate the metastable-only region from the regions of parameter space which include both stable and metastable vacua.}
	\label{fig:scenarios4}
\end{figure}

For scan 2, we find that the parameter points with stable minima are concentrated around small $\tan(\beta)< 0.9$ values (see Figure \ref{fig:scan2}). Note that for parameter points with $m_{h_3}>1000 \text{ GeV}$ and fixed $m^2_{12} = 3\cdot 10^4 \text{ GeV}^2$, the region of parameter points with stable vacua reduces to $\tan(\beta)< 0.7$. This is in agreement with the observation that the region with stable vacua disappears for all $\tan(\beta)$, for increasing $m^2_{12}$. We note, however, that several parameter points with metastable EW vacuum and several minima at $v_s=0$ exist for $\tan(\beta)> 1$. 

For scan 3, we fix $m_{h_3} = 600 \text{ GeV}$ and vary the other parameters. As shown in Figure \ref{fig:scan3}, we find that parameter region with only metastable vacua is concentrated in the region of low $m_{H^{\pm}}$ and $m_{A}$, as well as $( 2.5\cdot 10^4 <  m^2_{12} < 9 \cdot 10^4)  \text{ GeV}^2$. This region also corresponds to values $2 < \tan(\beta) < 3$.

Concerning scan 4, we find that the results depend on $m_{12}$ and $\tan\beta$. In this scan, parameter points with $(3 \times 10^4 <  m^2_{12} < 2 \cdot 10^5)  \text{ GeV}^2$ and low $\tan \beta <1$ are always metastable, while parameter points with higher $\tan \beta <3$ are metastable already at smaller $ (10^4 < m^2_{12} < 10^5) \text{ GeV}^2$ (See Figure \ref{subfig:scan4tanbetam12}). When extending the scan to higher values of $m^2_{12}$, we also find a region where stable and metastable vacua coexist, characterized by masses $m_A$ and $m_{H^\pm}$ of the order $\mathcal{O}(1 \text{ TeV})$. 

For scan 5, we find that regions of stability and metastability overlap with each other for all variables, and therefore, we cannot use this mechanism to rule out parameter regions, and the DW decay mechanism can only be used to rule out individual parameter points.
 
Summarizing our results, we find that for this particular scenario, the variable $m^2_{12}$, and to a lesser extent $\tan(\beta)$, play a crucial role in determining parameter regions that are only metastable. However, we found that symmetry restored parameter points, as well as parameter points with possible symmetry non-restoration (according to the Arnold-Espinosa resummation scheme), overlap. In order to be able to conclusively rule out those parameter regions using DW induced vacuum decay, this possibility of symmetry non-restoration needs to be addressed in more detail, which is beyond the scope of the current work.

\subsubsection{Scan including experimental constraints}\label{experimental}
\begin{figure}[h]
     \centering
    \subfloat[]{\includegraphics[width=0.5\columnwidth]{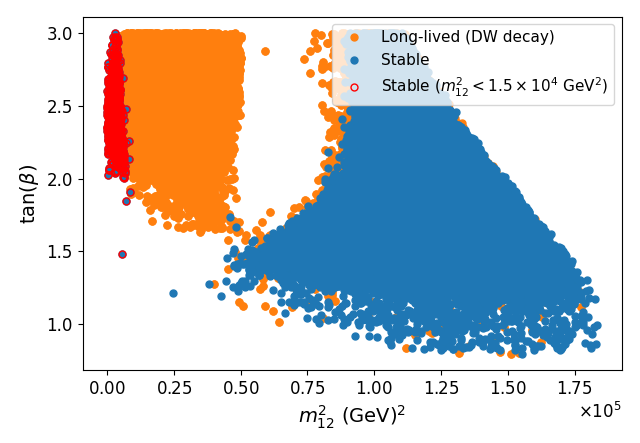}
         \label{subfig:m12tanbeta1600}}
    \subfloat[]{\includegraphics[width=0.5\columnwidth]{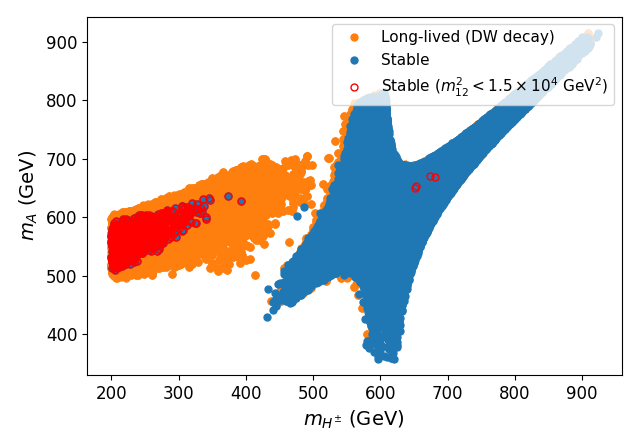}
         \label{subfig:mhmc1600}} \\
\caption{Parameter scan for $m_{h_3} = 600 \text{ GeV}$ after applying all experimental constraints. The parameter points shown in black circles have a stable EW vacuum but with very low values of $m^2_{12}<1.5 \cdot 10^4 \text{ GeV}^2$.} 
\label{fig:scenario1exp}
\end{figure}
\begin{figure}[h]
     \centering
     \begin{subfigure}[b]{0.32\textwidth}
         \centering
         \includegraphics[width=\textwidth]{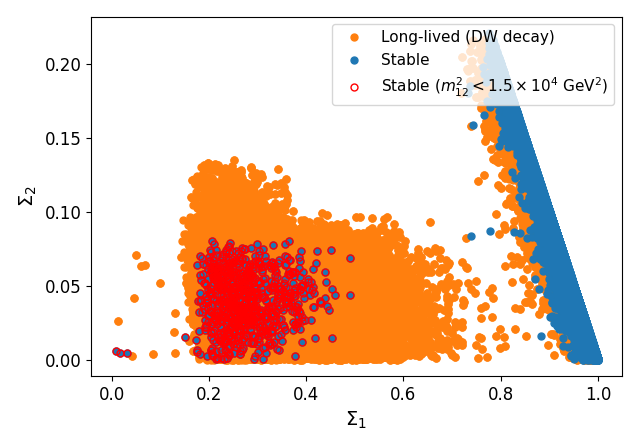}
       \subcaption{}  \label{subfig:s1s21400}
     \end{subfigure}
     \hfill
     \begin{subfigure}[b]{0.32\textwidth}
         \centering
         \includegraphics[width=\textwidth]{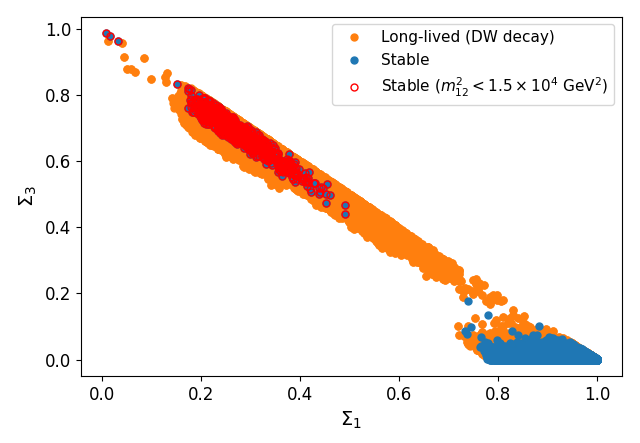}
        \subcaption{} \label{subfig:s1s31400}
     \end{subfigure}
     \begin{subfigure}[b]{0.32\textwidth}
         \centering
         \includegraphics[width=\textwidth]{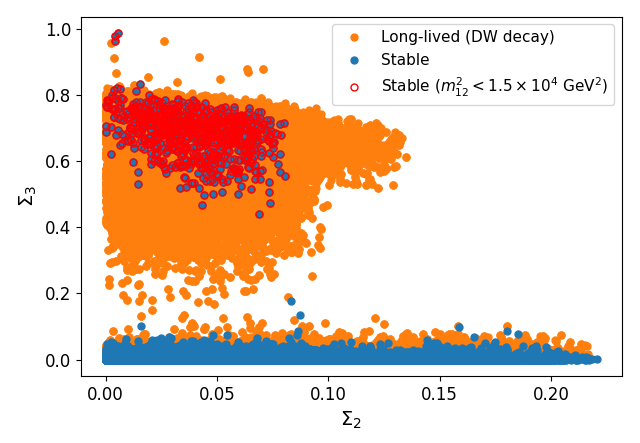}
        \subcaption{} \label{subfig:s2s31400}
     \end{subfigure}
   
\caption{Scenario 1: Stability of the EW minimum for different parameter scans as a function of $\Sigma_i$, the singlet admixture in the CP-even Higgs bosons. The parameter points shown in black circles have a stable EW vacuum but with very low values of $m^2_{12}<15 \cdot 10^3 \text{ GeV}^2$.} 
\label{fig:scenario1anglesexp}
\end{figure}
We now generate parameter points with \texttt{ScannerS}, taking into account experimental constraints such as collider searches, flavor constraints, and electroweak precision measurements. We focus on a parameter scan with $m_{h_3} = 600 \text{ GeV}$.   

We find that the region where only metastable vacua are obtained is independent on $v_s$. Again, one can use the variable $m^2_{12}$ to differentiate between stable and metastable regions. We find that lower values lead to regions with mostly metastable EW minima, while parameter points with higher values of $m^2_{12}$ lead to both stable and metastable EW minima. However, for very small $m^2_{12}<1.5 \cdot 10^4 \text{ GeV}^2$, we can also obtain parameter points with stable EW vacua (marked with red circles). Therefore, one can use the mechanism of vacuum decay via domain walls to exclude lower values of $m^2_{12}$ in this scenario up to $m^2_{12} = 1.5 \cdot 10^4 \text{ GeV}^2$.

We also find a correlation between $m_{H^\pm}$ and $m_A$ and the regions featuring mostly metastable EW minima as shown in Figure \ref{subfig:mhmc1600}. This is the case for mostly lower values for $m_{H^{\pm}}$, and $m_A$ and $m^2_{12}>1.5 \cdot 10^4 \text{ GeV}^2$.

Finally, one can use the singlet admixture $\Sigma_i = R^2_{i3}$ in order to differentiate between regions of EW minimum metastability and stability. We find that parameter points where the singlet admixture in $h_3$ is higher lead to regions with metastable EW vacua only. 

\subsection{Scenario 2: SM Higgs as the lightest CP-even Higgs boson}

We now consider the case when the 125 GeV SM-like Higgs is the lightest CP-even Higgs boson. We fix the second CP-even Higgs mass to $m_{h_2} = 400 \text{ GeV}$, and allow $m_{h_3} \in \{600, 800, 1200\} \text{ GeV}$. We keep the same scan ranges for all the other model parameters as shown in Table \ref{tab:scans}.

\begin{figure}[h]
	\centering
	\subfloat[$m_{h_3} = 600 \text{ GeV}$]{\includegraphics[width=0.5\columnwidth]{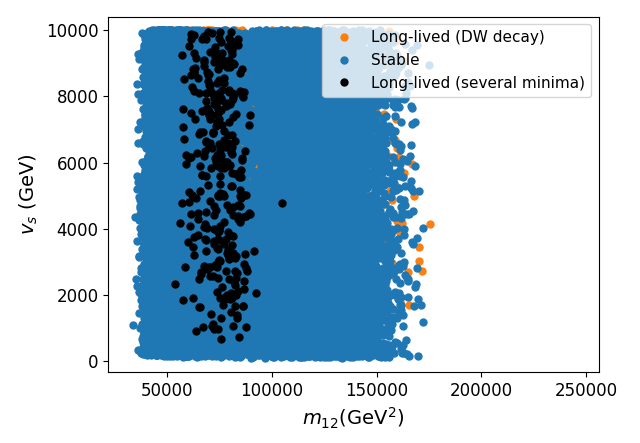}
		\label{subfig:m12vs2600}}
	\subfloat[$m_{h_3} = 600 \text{ GeV}$]{\includegraphics[width=0.5\columnwidth]{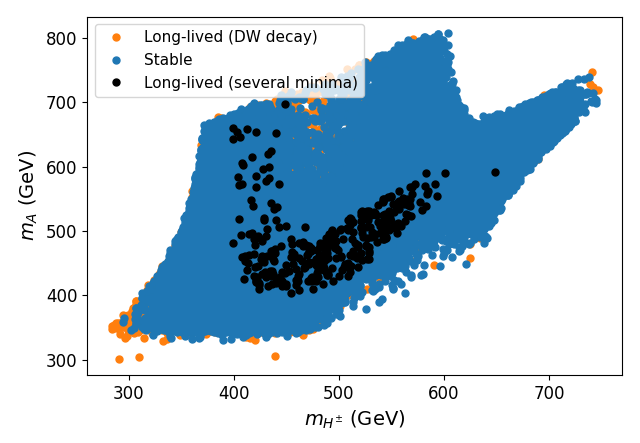}
		\label{subfig:mhmc2600}} \\
	\subfloat[$m_{h_3} = 800 \text{ GeV}$]{\includegraphics[width=0.5\columnwidth]{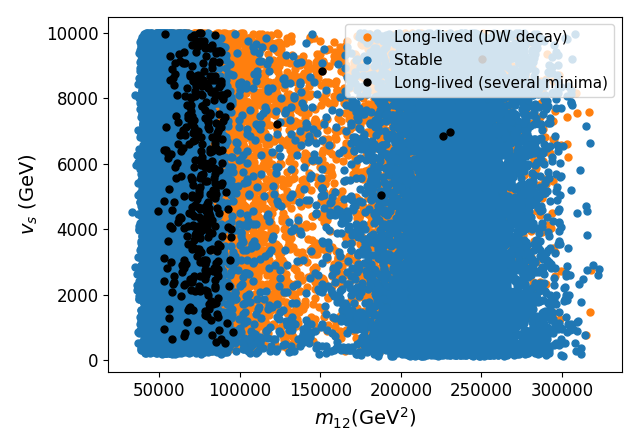}
		\label{subfig:m12vs2800}} 
	\subfloat[$m_{h_3} = 800 \text{ GeV}$]{\includegraphics[width=0.5\columnwidth]{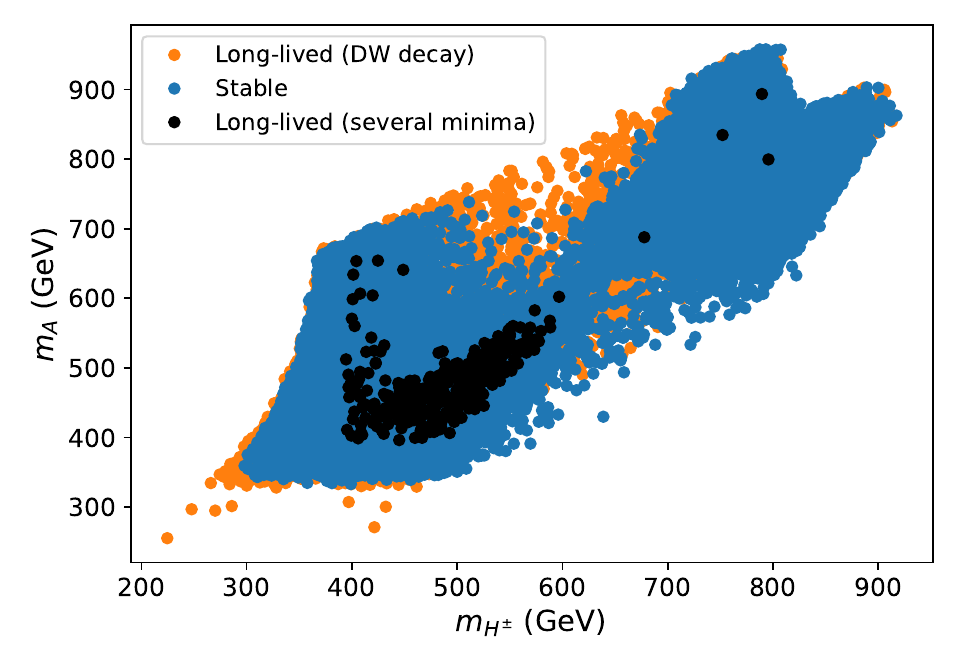}
		\label{subfig:mhmc2800}} \\
	\subfloat[$m_{h_3} = 1200 \text{ GeV}$]{\includegraphics[width=0.5\columnwidth]{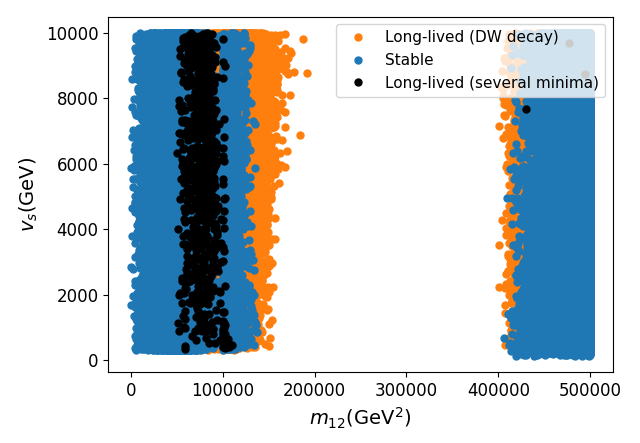}
		\label{subfig:m12vs21200}}
	\subfloat[$m_{h_3} = 1200 \text{ GeV}$]{\includegraphics[width=0.5\columnwidth]{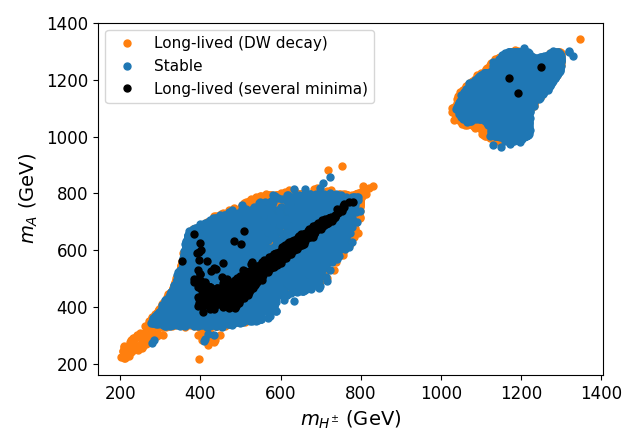}
		\label{subfig:mhmc21200}}
	
	\caption{Scenario 2: Stability of the EW minimum for different parameter scans as a function of the singlet VEV $v_s$ and $Z_2$ breaking term $m_{12}$. (a) and (b) correspond to $m_{h_3} = 600\text{ GeV}$, (c) and (d) correspond to $m_{h_3} = 800 \text{ GeV}$, and (e) and (f) correspond to $m_{h_3}=1200\text{ GeV}$. } 
	\label{fig:scenario2masses}
\end{figure}

\begin{figure}[h]
	\centering
	\subfloat[$m_{h_3} = 600 \text{ GeV}$]{\includegraphics[width=0.32\columnwidth]{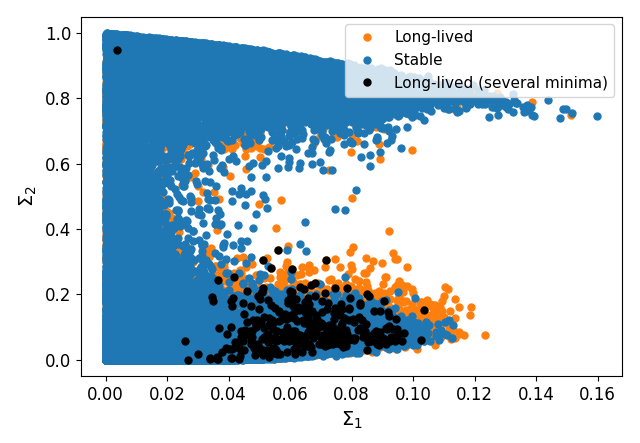}
		\label{subfig:s1s2600}}
	\subfloat[$m_{h_3} = 600 \text{ GeV}$]{\includegraphics[width=0.32\columnwidth]{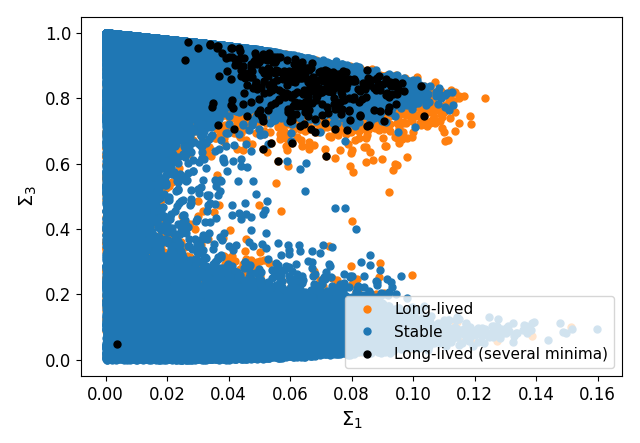}
		\label{subfig:s1s32600}}
	\subfloat[$m_{h_3} = 600 \text{ GeV}$]{\includegraphics[width=0.32\columnwidth]{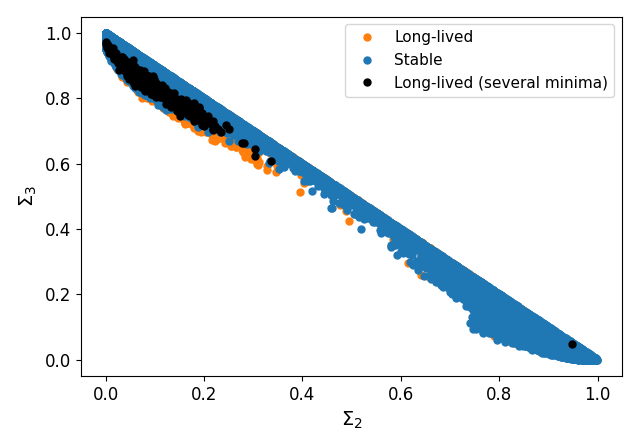}
		\label{subfig:s2s32300}} \\
	\subfloat[$m_{h_3} = 800 \text{ GeV}$]{\includegraphics[width=0.32\columnwidth]{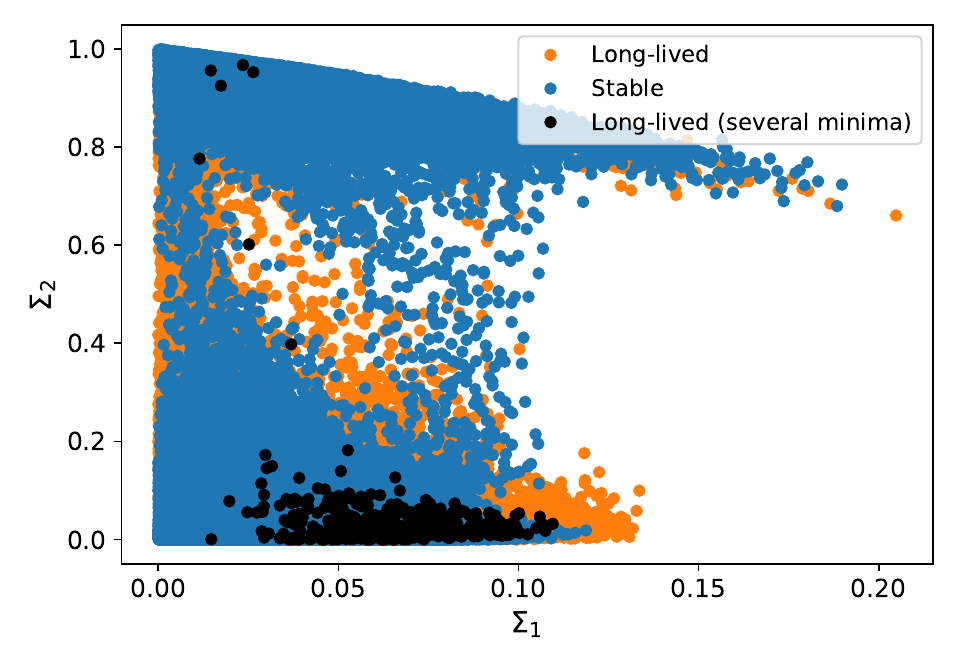}
		\label{subfig:s1s22800}}
	\subfloat[$m_{h_3} = 800 \text{ GeV}$]{\includegraphics[width=0.32\columnwidth]{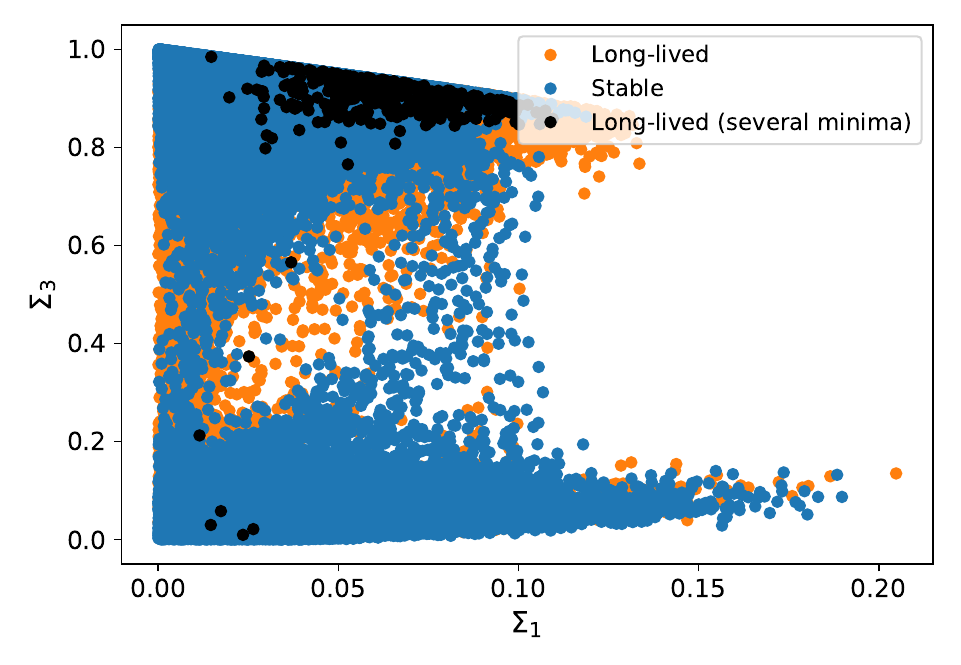}
		\label{subfig:s1s32800}}
	\subfloat[$m_{h_3} = 800 \text{ GeV}$]{\includegraphics[width=0.32\columnwidth]{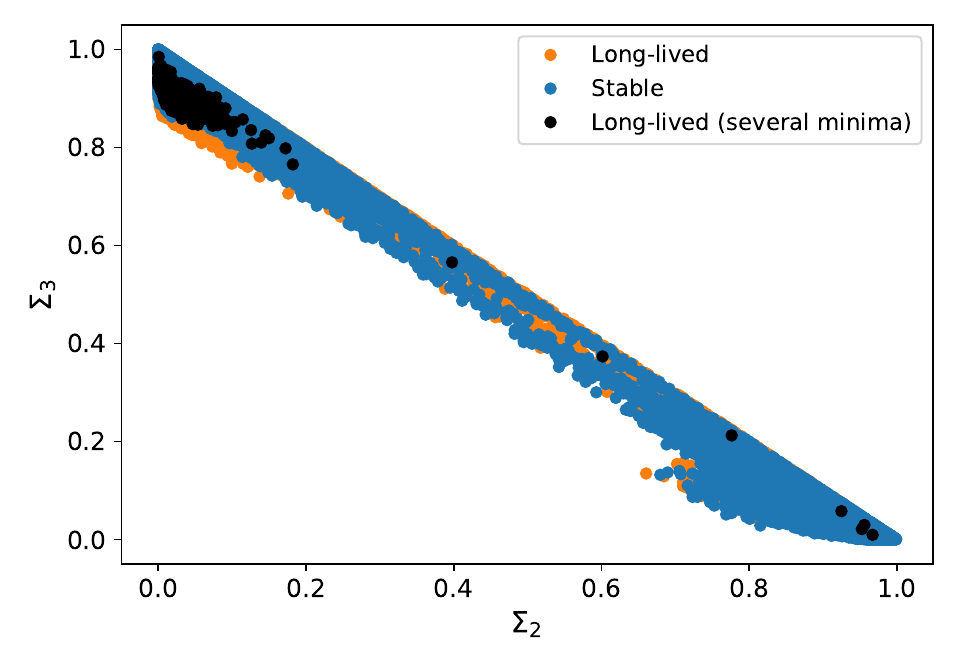}
		\label{subfig:s2s32800}} \\
	\subfloat[$m_{h_3} = 1200 \text{ GeV}$]{\includegraphics[width=0.32\columnwidth]{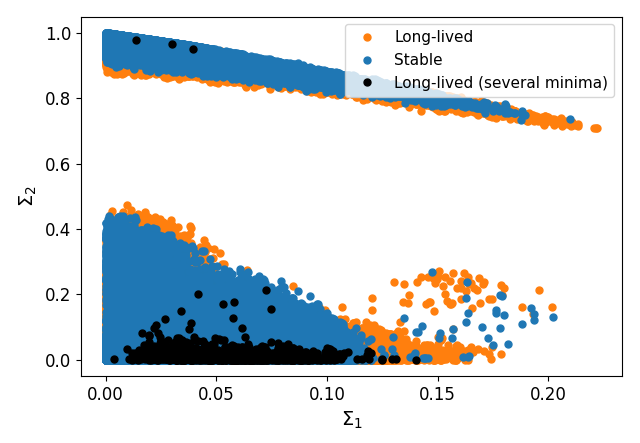}
		\label{subfig:s1s221200}}
	\subfloat[$m_{h_3} = 1200 \text{ GeV}$]{\includegraphics[width=0.32\columnwidth]{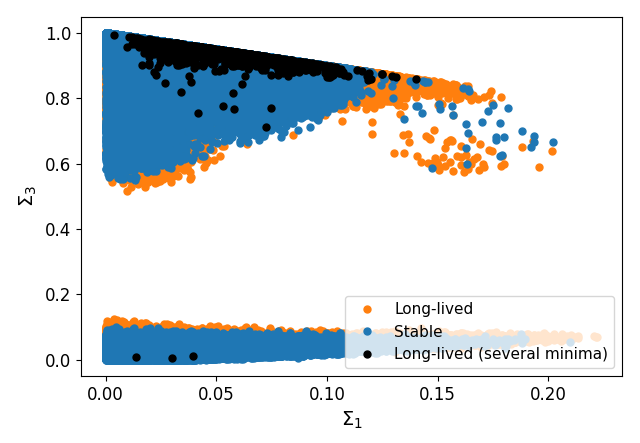}
		\label{subfig:s1s321200}}
	\subfloat[$m_{h_3} = 1200 \text{ GeV}$]{\includegraphics[width=0.32\columnwidth]{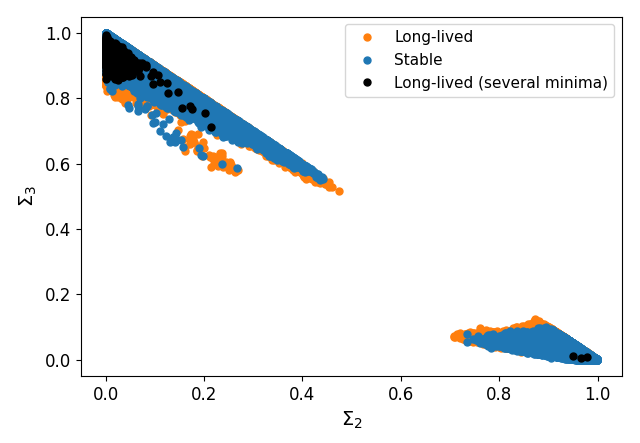}
		\label{subfig:s2s321200}}
	\caption{Scenario 2: Stability of the EW minimum for different parameter scans as a function of $\Sigma_i = R^2_{i3}$, the singlet admixture in the CP-even Higgs bosons. Upper row corresponds to the scan with $m_{h_3} = 600 \text{ GeV}$, middle row corresponds to $m_{h_3} = 800 \text{ GeV}$, and lower row to $m_{h_3} = 1200 \text{ GeV}$. } 
	\label{fig:scenario2angles}
\end{figure}

The results are shown in Figures \ref{fig:scenario2masses} and \ref{fig:scenario2angles}. For this case, the distinction between parameter point regions with solely metastable minima and regions with both stable and metastable minima is not possible, and one cannot use the decay of the EW vacuum induced by the domain walls in order to systematically rule out specific parameter regions like in scenario 1.

We obtain several parameter points featuring at least two minima for the potential at $v_s = 0 \text{ GeV}$ (shown in black in Figures \ref{fig:scenario2masses} and \ref{fig:scenario2angles}). In all three scans, these points were concentrated around $ (5 \cdot 10^4 < m^2_{12} < 10^5) \text{ GeV}^2 $, $\tan(\beta) > 1.5$, as well as around the same range of the masses for $m_{H^{\pm},A}$ and the singlet admixture, which shows that this effect is more pronounced for parameter points with a large singlet admixture in the heavier CP-even Higgs. Note that the singlet admixture in the SM-like Higgs boson $h_1$ has to be quite low in these scans since larger values are already ruled out by Higgs measurement and searches at the LHC \cite{Bechtle:2020pkv, Bechtle:2020uwn,Bechtle:2015pma}. It is therefore possible, for these parameter points, that the global minimum is not nucleated inside the domain wall and instead, the scalar fields converge to the intermediate minimum. For those parameter points, we explicitly calculate the real-time evolution (\ref{eq:eom}) of the DW solution to determine the fate of the metastable EW minimum using $\hat{d} \approx 0$. We find that all these parameter points roll over to the global minimum, leading to the decay of the long-lived EW vacuum. 
One cannot, however, conclude that any parameter point with multiple minima will necessarily experience vacuum decay via domain walls, and the possibility that the field configuration gets trapped inside the intermediate vacuum for some other parameter points is still possible. In such a case, the tunneling rate between the trapped field configuration and a field configuration where the global vacuum is nucleated inside the wall has to be calculated. We leave the investigation of this possibility to a future work.

\section{Summary and conclusions}
\label{section6}
In this work, we investigated the decay induced by domain walls of long-lived EW vacua in the N2HDM. This mechanism of vacuum decay can be used in order to rule out a large number of viable parameter points in several scenarios. 
In the standard N2HDM, where the model is invariant under the $Z'_2$ symmetry, the formation of domain walls related to the scalar singlet leads to the possibility that the 2HDM part of the potential inside the wall can be very different with respect to that potential outside of the wall. For parameter points where the global minimum has a vanishing singlet VEV $v_s$, the field configuration inside the domain wall can roll over to the global minimum. Since the potential of the global minimum is lower than the EW metastable minimum outside the wall, the global vacuum is nucleated inside the wall and then quickly expands outside, leading to the total decay of the EW vacuum. Those parameter points are thus not allowed since the masses of all particles will have very different values from the ones observed. 

We checked that the mechanism of vacuum decay occurs for all types of vacua in the N2HDM: neutral, electrically charged as well as CP-violating vacua. We also showed that one can use this mechanism in certain scenarios to rule out some regions of parameter points. In particular, we found that in the case of a possible 95-GeV Higgs boson, the region of metastability depends heavily on the values of $m_{h_3}$, $m^2_{12}$, and $\tan(\beta)$.

One important aspect that we found in our work is the possibility that, inside the wall, several minima in the 2HDM part of the potential can coexist alongside the global minimum. This observation prevents the direct ruling out of a parameter point of the model based only on the metastability of the EW vacuum and its decay induced by the domain wall. In such cases, one needs to explicitly compute the real-time evolution of the scalar field configuration inside the domain wall to determine whether the global vacuum will be nucleated inside the wall. Even though several parameter points featured such a case of multiple minima, we didn't find regions where the field configuration inside the wall is trapped in the intermediate vacuum. If a parameter point leads to such behavior, one then needs to calculate the tunneling rate between the trapped field configuration and the field configuration where the global vacuum is nucleated inside the wall. These calculations are non-trivial and numerically time-consuming. We therefore leave the discussion of quantum tunneling induced by the domain wall field configuration to future work.

We showed that many parameter points of the $Z'_2$ symmetric N2HDM would be ruled out due to EW vacuum decay induced by domain walls. This mechanism provides strong constraints for this model by ruling out parameter points that lead to long-lived metastable EW vacua where the global vacuum has a vanishing singlet VEV. There are, however, possible ways to circumvent this constraint. First, one could break the $Z'_2$ symmetry by introducing large symmetry-breaking terms in order to avoid the formation of domain walls altogether. This, however, can considerably alter the phenomenology of the N2HDM. Another way is to only choose parameter points where the $Z'_2$ symmetry is not spontaneously broken and the EW vacuum has $v_s=0$. Such a choice is also motivated by the possibility of generating DM candidates \cite{Glaus:2022rdc, Engeln:2020fld}. Alternatively, one could also use parameter points that feature $Z'_2$ symmetry non-restoration, which avoids the formation of the domain walls in the early universe. This final solution requires, however, a detailed and careful computation of the daisy resummation needed for the thermal potential in order to obtain reliable results.  

\appendix

\acknowledgments
We thank Simone Blasi and the DESY 2HDM working group for helpful discussions. This work is funded by the Deutsche Forschungsgemeinschaft (DFG) through Germany’s Excellence Strategy – EXC 2121 “Quantum Universe” — 390833306. Figures presented in this work were generated using \texttt{MatPlotLib} \cite{Hunter:2007} and \texttt{NumPy} \cite{2020NumPy-Array}.
\appendix

\section{Constraints}\label{appendixconstraints}
We briefly describe the constraints imposed on the parameter scans generated in our work.
\subsection{Theoretical Constraints}
We include constraints based on the theory consistency of the model, such as unitarity, a well-behaved potential bounded from below, and the need for an EW minimum that is either stable or metastable with a lifetime larger than the age of the universe. 
\subsubsection{Perturbative unitarity}
In quantum field theory, the scattering matrix $S$ has to be unitary, $S^{\dagger}S = I$, in order to keep a conserved transition probability \cite{Schwartz_2013}. 

When extending the Higgs sector with extra multiplets, it is crucial to make sure that the scalar contributions cancel the divergent parts in the scattering amplitude of longitudinal $W$ bosons \cite{Gunion:1989we, Branco:2011iw}. For n extra Higgs doublets/singlets, this translates into the condition \cite{Gunion:1989we}:
\begin{equation}
	\sum_{i}^{n} g^2_{h_iVV} = g^2_{H_{SM}VV},
\end{equation}
where $g^2_{h_iVV}$ denotes the coupling of the extra Higgs bosons to the weak gauge bosons, and $g_{H_{SM}VV}$ denotes the coupling of the SM-like Higgs to the gauge bosons. 
One also needs to make sure that the remaining finite contributions in the scattering of the longitudinal $W$ bosons fulfill the condition of perturbative unitarity. The same procedure can be done using the Goldstone boson scattering according to the equivalence theorem \cite{Gunion:1989we}. In models with several scalar doublets, one can construct a matrix $(a_0)_{ij}$ corresponding to the different scattering amplitudes of the Goldstone bosons $(\mathcal{M}_{2\rightarrow 2})_{ij}$ \cite{Branco:2011iw}.
The bound on perturbative unitarity can be expressed as a bound on the eigenvalues of this matrix:
\begin{equation}
	\abs{\mathcal{M}^i_{2 \rightarrow 2}} < 8\pi.
\end{equation}  
These eigenvalues are expressed in terms of the quartic couplings of the scalar potential.
For the N2HDM, this translates into the following constraints \cite{Muhlleitner:2016mzt}:
\begin{align}
&	\abs{\lambda_3 \pm \lambda_4} < 8\pi, \\
&   \abs{\lambda_3 \pm \lambda_5 } < 8\pi, \\
& \abs{\lambda_3 + 2\lambda_4 \pm 3\lambda_5} < 8\pi, \\
& \abs{\frac{1}{2}\left( \lambda_1 + \lambda_2 \pm \sqrt{(\lambda_1-\lambda_2)^2+ 4\lambda^2_4}   \right)} < 8\pi, \\
& \abs{\frac{1}{2}\left( \lambda_1 + \lambda_2 \pm \sqrt{(\lambda_1-\lambda_2)^2+ 4\lambda^2_5}   \right)} < 8\pi, \\
& \abs{\frac{1}{2}\left( 3\lambda_1 + 3\lambda_2 \pm \sqrt{9(\lambda_1-\lambda_2)^2+ 4(2\lambda_3+\lambda_4)^2}   \right)} < 8\pi, \\
&	\abs{\lambda_7} < 8\pi, \\
&   \abs{\lambda_8} < 8\pi, \\
& \frac{1}{2}\abs{a_{1,2,3}} < 8\pi,
\end{align}
where $a_{1,2,3}$ are the roots of the polynomial:
\begin{align}
\notag & 4(-27\lambda_1\lambda_2\lambda_6 + 12\lambda^2_3\lambda_6 + 12\lambda_3\lambda_4\lambda_6+ 3\lambda^2_4\lambda_6 + 6\lambda_2\lambda^2_7 - 8\lambda_3\lambda_7\lambda_8 -4\lambda_4\lambda_7\lambda_8 \\ \notag & + 6\lambda_1\lambda^2_8) + x(36\lambda_1\lambda_2 - 16\lambda^2_3 - 16\lambda_3\lambda_4 - 4\lambda^2_4 + 18\lambda_1\lambda_6  - 4\lambda^2_7 - 4\lambda^2_8) \\ & - 3x^2(\lambda_6 - 2(\lambda_1+ \lambda_2)) + x^3.
\end{align}

\subsubsection{Boundedness from below}
In the SM, the condition to obtain a non-zero and finite minimum for the potential of the Higgs is to ensure that the quadratic term $\mu^2 < 0$ is negative, and $\lambda > 0$. This ensures that there are no directions in the field where the potential falls into $-\infty$, making it unbounded and causing an instability in the potential. 

In extended Higgs sectors, several new terms are added to the scalar potential, making the potential much more complex, and therefore, it is important to make sure that no direction in the multidimensional space of the scalar fields can lead to the potential falling into $-\infty$.

For the N2HDM, the region of parameter space that is allowed is the union of the sets $\Omega_1 \cup \Omega_2$ where \cite{Muhlleitner:2016mzt}:
\begin{align}
\notag  \Omega_1 = & \biggl\{  \lambda_{1,2,6} > 0;\ \sqrt{\lambda_1\lambda_6}+\lambda_7 > 0;\ \sqrt{\lambda_2\lambda_6} + \lambda_8; \ \sqrt{\lambda_1\lambda_2} + \lambda_3 + D > 0; \\  & \lambda_7 + \sqrt{\lambda_1/\lambda_2}\lambda_8 \geq 0   \biggr\}, \\
\notag \Omega_2 = & \biggl\{  \lambda_{1,2,6} > 0;\ \sqrt{\lambda_1\lambda_6} \geq \lambda_8 > -\sqrt{\lambda_2\lambda_6};\ \sqrt{\lambda_1\lambda_6} > -\lambda_7 \geq \sqrt{\lambda_1/\lambda_8}; \\ & \sqrt{(\lambda^2_7-\lambda_1\lambda_6)(\lambda^2_8-\lambda_2\lambda_6)} > \lambda_7\lambda_8 -(D+\lambda_3)\lambda_6  \biggr\}, 
\end{align}
where $D = min(\lambda_4 - \abs{\lambda_5},0)$.
\subsubsection{Vacuum stability}

If the EW minimum is only local, one needs to calculate the transition rate to the global one. This rate $\Gamma$ per unit volume $V$ is given by \cite{Ferreira:2019iqb, PhysRevD.15.2929}:
\begin{equation}
	\dfrac{\Gamma}{V} = Ke^{-B},
\end{equation}
where $K$ is a dimensionful constant related to the electroweak scale and $B$ is the bounce action describing a scalar field configuration between the electroweak to the global vacuum. It was found in \cite{Ferreira:2019iqb} that for $B \lesssim 390$, the EW vacuum is unstable and therefore those parameter points will be ruled out. In case $B>440$, the electroweak vacuum is metastable with a lifetime larger than the age of the universe, making the EW vacuum phenomenologically viable.

\subsubsection{Symmetry restoration in the early universe}
In order to produce the $Z'_2$ domain walls, the $Z'_2$ symmetry needs to be spontaneously broken in the early universe. Therefore, this symmetry should be restored at high temperatures. In case this symmetry is not restored at high temperatures (i.e. the minimum of the potential has a non-zero $v_s$), the domain walls would not have been formed in the early universe, and one wouldn't be able to use them to induce EW vacuum decay.

The phenomenon of symmetry non-restoration in the N2HDM was already extensively studied in \cite{Biekotter:2021ysx}. We here use the analytical formulas provided in \cite{Biekotter:2021ysx} in order to verify the possibility of symmetry non-restoration.

\noindent
The temperature-dependent potential of the N2HDM has the form: 
\begin{equation}
  \notag  V_{N2HDM}(T, \Phi_1, \Phi_2, \Phi_s) = m^2_{11}(T) |\Phi_1|^2 + m^2_{22}(T) |\Phi_2|^2 + m^2_S(T) \Phi^2_s +  m^2_{12}\Phi_1\Phi_2  + ... ,
\end{equation}
where the parameters of the potential are temperature-dependent. It is therefore possible that for a non-zero temperature $T \neq 0$, the effective mass terms turn positive, shifting the minimum of the potential to be at the origin of field space \textcolor{black}{$(\Phi_1, \Phi_2, \Phi_s)_T = (0, 0, 0)$} and therefore leading to the restoration of the symmetry. To investigate this, one can use both an analytical as well as a numerical approach. The analytical approach discussed in \cite{Biekotter:2021ysx} calculates the Hessian matrix of the potential at its origin (0,0,0) at high temperatures T, which gives us information about the curvature of the potential around the origin. This involves calculating the principal minors of the Hessian matrix $H^0_{i,j} = \partial^2V/\partial\phi_i\partial\phi_j|_{(0,0,0)}$. One can then define the quantities\footnote{These coefficients also incorporate terms from the daisy resummation of infrared-divergent contributions in the thermal potential. The authors of \cite{Biekotter:2021ysx} use the Arnold-Espinosa method \cite{Arnold:1992rz} for the derivation of the coefficients $c_{ii}$. } $c_{ii} \equiv
\lim\limits_{\,T\rightarrow\infty}\,
H_{ii}^{0}/ T^2 $ \cite{Biekotter:2021ysx}:
\begin{align}
c_{11} & \simeq
-0.025+c_{1}-\frac{1}{2\pi}\left(\frac{3}{2}\lambda_{1}\sqrt{c_1}+\lambda_{3}\sqrt{c_{2}}+\frac{1}{2}\lambda_{4}\sqrt{c_{2}}+\frac{1}{4}\lambda_{7}\sqrt{c_{3}}\right)
\ ,
\label{coeff_1}
\\
c_{22} & \simeq
-0.025+c_{2}-\frac{1}{2\pi}\left(\frac{3}{2}\lambda_{2}\sqrt{c_2}+\lambda_{3}\sqrt{c_{1}}+\frac{1}{2}\lambda_{4}\sqrt{c_{1}}+\frac{1}{4}\lambda_{8}\sqrt{c_{3}}\right)
\ ,
\label{coeff_2}
\\
c_{33} & =
c_{3}-\frac{1}{2\pi}\left(\lambda_{7}\sqrt{c_1}+\lambda_{8}\sqrt{c_{2}}+\frac{3}{4}\lambda_{6}\sqrt{c_{3}}\right)\, ,
\label{coeff_3}
\end{align}
where the coefficients $c_{i}$ are defined as \cite{Biekotter:2021ysx}:
\begin{align}
\label{c1} 
c_{1} &=
\frac{1}{16}({g'}^{2}+3g^{2})+\frac{\lambda_{1}}{4}+\frac{\lambda_{3}}{6}+\frac{\lambda_{4}}{12}+\frac{\lambda_{7}}{24}\,
,\\
\label{c2}
c_{2} &=
\frac{1}{16}({g'}^{2}+3g^{2})+\frac{\lambda_{2}}{4}+\frac{\lambda_{3}}{6}+\frac{\lambda_{4}}{12}+\frac{\lambda_{8}}{24}+\frac{1}{4}y_{t}^{2}
\, ,\\
\label{c3}
c_{3} &= \frac{1}{6}(\lambda_{7}+\lambda_{8})+\frac{1}{8}\lambda_{6} \, ,
\end{align} 
with g and $g'$ denoting the weak gauge couplings and $y_t$ the Yukawa coupling to the top quark. For positive $c_{11}$ and $c_{22}$, the electroweak symmetry is restored at high T and in case $c_{33}>0$, the $Z'_2$ symmetry is restored too. In this work, we focus on the restoration of the $Z'_2$ symmetry at higher temperatures to ensure the formation of the singlet domain walls. In case when $c_{11,22} < 0$, the doublets have a temperature-dependent VEV, and the required Hessian matrix has to be evaluated at $(v_1(T), v_2(T), 0)$ to reliably determine whether the $Z'_2$ symmetry is restored. Such a calculation is more complicated and can only be done numerically for the N2HDM.

This constraint was included in our implementation for \texttt{ScannerS} \cite{Coimbra:2013qq, Ferreira:2014dya, Muhlleitner:2016mzt, Costa:2015llh, Muhlleitner:2020wwk} in order to verify if the generated parameter points for our scans feature the EW and $Z'_2$ symmetries restoration at some stage in the early universe. Note that the expressions for $c_{11}$, $c_{22}$ and $c_{33}$ derived in \cite{Biekotter:2021ysx} use the Arnold-Espinosa resummation scheme \cite{Arnold:1992rz}. It is known that different resummation schemes can lead to different outcomes for the thermal evolution of the scalar fields in the early universe (see e.g. \cite{Bittar:2025lcr} and page 23 in \cite{Biekotter:2021ysx}). A detailed inclusion of these aspects is beyond the scope of this work, and therefore, we restrict our analyses and use these formulas in the framework of the Arnold-Espinosa method only as an attempt to verify the restoration of the $Z'_2$ symmetry in the early universe.  
\subsection{Experimental Constraints}

\subsubsection{Collider Constraints}
The Higgs boson was the most recent fundamental particle, discovered in 2012 at the LHC. Until now, the experimental searches in colliders have only put exclusion limits on viable parameter regions of BSM Higgs models.

There are several tools used in order to test the viability of parameter points in extended Higgs sectors with the exclusion limits provided by experiments at the LHC, such as \texttt{HiggsBounds} \cite{Bechtle:2008jh, Bechtle:2011sb, Bechtle:2012lvg, Bechtle:2013wla, Bechtle:2013xfa, Bechtle:2015pma, Bechtle:2020pkv}. This tool extracts the experimental upper bounds from several BSM Higgs searches at the LHC, LEP, and the Tevatron and compares the prediction for the cross section times the branching ratio of a given parameter point with the observed upper limit at the experiment. If this ratio is larger than one, then the parameter point can be excluded since it would lead to a larger number of events than the maximum number allowed by experiments. If the ratio is smaller than one, then the parameter point is still allowed since it leads to a prediction that is still in agreement with the experimental limits. 

Another important experimental constraint is the fact that, so far, the properties of the observed Higgs boson at the LHC mostly agree with SM predictions within the experimental uncertainties. Therefore, any BSM Higgs model needs to predict a SM-like Higgs boson in its particle spectrum with properties (such as decay, production, and couplings) similar to those predicted by the SM. Since the LHC measurements of the 125 GeV scalar particle still allow some deviations from the SM prediction, one can compare the properties of the SM-like Higgs boson in the BSM model within the experimental uncertainties. One tool dedicated to such analyses is \texttt{HiggsSignals} \cite{Bechtle:2013xfa, Stal:2013hwa, Bechtle:2014ewa, Bechtle:2020uwn}. 
\subsubsection{Flavor Constraints}
Important other constraints to constrain the BSM Higgs sector originate from flavor physics. We mentioned earlier that the $Z_2$ symmetry introduced in the 2HDM is necessary in order to forbid FCNC at tree level. However, loop corrections involving BSM particles can also induce FCNC. In the 2HDM and N2HDM, these contributions are mainly produced by the charged Higgs bosons $H^{\pm}$ and can be relevant in low-energy B-meson decays \cite{Haller:2018nnx}.
\subsubsection{Electroweak Precision Observables}
For a model with n scalar multiplets, the parameter $\rho = \dfrac{\sum^n_{i=1}[I_i(I_i+1)-\frac{1}{4}Y^2_i]v_i}{\sum^{n}_{i=1}\frac{1}{2}Y^2_iv_i}$ for any Higgs sector has to be equal to one at tree level \cite{Gunion:1989we}, where $I_i$ corresponds to the weak isospin of the i-th scalar multiplet, $Y_i$ to its weak hypercharge, and $v_i$ to its VEV. The experimentally measured value is $\rho = (1.00031 \pm 0.00019)$ \cite{PhysRevD.110.030001}. The small deviations from one are due to loop contributions to the gauge bosons' self-energies. If BSM Higgs models contribute to these self-energies, then one needs to make sure that the contributions do not exceed the experimentally measured value. In BSM Higgs models, these contributions often can be conveniently incorporated in the framework of the oblique parameters $S$, $T$, and $U$ defined by \cite{Grimus:2007if}:
\begin{align}
&	\alpha(m_Z)T = \dfrac{\Pi_{WW}(0)}{m^2_{W}} - \dfrac{\Pi_{ZZ}(0)}{m^2_{Z}}, \\
& \notag	\dfrac{\alpha(m_Z)}{4\sin^2(\theta_W)\cos^2(\theta_W)}S = \dfrac{\Pi_{ZZ}(m^2_Z)-\Pi_{ZZ}(0)}{m^2_Z} - \dfrac{\cos^2(\theta_W)-\sin^2(\theta_W)}{\sin(\theta_W)\cos(\theta_W)}\dfrac{\Pi_{Z\gamma}(m^2_Z)}{m^2_Z} \\ & - \dfrac{\Pi_{\gamma\gamma}(m^2_Z)}{m^2_Z}, \\
& \dfrac{\alpha(m_Z)}{4\sin^2(\theta_W)}(S+U) = \dfrac{\Pi_{WW}(m^2_W)-\Pi_{WW}(0)}{m^2_W} - \dfrac{\cos(\theta_W)}{\sin(\theta_W)}\dfrac{\Pi_{Z\gamma}(m^2_Z)}{m^2_Z} - \dfrac{\Pi_{\gamma\gamma}(m^2_Z)}{m^2_Z}, 
\end{align}
where $\Pi_{ij}$ denotes the one-loop self energies of the gauge bosons, and $\alpha(m_Z)$ denotes the weak coupling constant at the $Z$ scale. The experimental limits for these parameters are given by \cite{PhysRevD.110.030001}:
\begin{align} 
	S = -0.04 \pm 0.1, &&  T = 0.01 \pm 0.1, && 
	U = -0.01 \pm  0.09.
\end{align}	

All these theoretical and experimental limits were imposed in our scans (unless claimed otherwise) using the public code \texttt{ScannerS} \cite{Muhlleitner:2020wwk}, which generates random parameter points in a given range of parameter variables that fulfill all the listed constraints.

\bibliography{references.bib}

\end{document}